\documentclass[aps,prb,showpacs,superscriptaddress,longbibliography]{revtex4-2} \usepackage{physics}
\usepackage{graphicx}
\usepackage{bm}
\usepackage{hyperref}
\usepackage[normalem]{ulem}
\usepackage{bbold}

\usepackage{comment}

\usepackage{siunitx}
\usepackage{layouts}
\usepackage[dvipsnames]{xcolor}
\usepackage{xurl}

\setcitestyle{super,open={},close={}}

\newcommand{\gone}{V_\mathrm{G1}}
\newcommand{\gtwo}{V_\mathrm{G2}}
\newcommand{\gthree}{V_\mathrm{G3}}
\newcommand{\vsd}{V_\mathrm{SD}}
\newcommand{\gs}{g_\mathrm{s}}
\newcommand{\gc}{g_\mathrm{c}}
\newcommand{\tsc}{t_\mathrm{sc}}
\newcommand{\tsf}{t_\mathrm{sf}}
\newcommand{\lso}{\ell_\mathrm{so}}
\newcommand{\nso}{\mathbf{n}_\mathrm{so}}
\newcommand{\gt}{\mathsf{g}}
\newcommand{\gp}{\tilde{\gt}}

\hypersetup{
    colorlinks,
    linkcolor={blue!50!black},
    citecolor={blue!50!black},
    urlcolor={blue!80!black}
}

\begin{document}
\title{Strong coupling between a photon and a hole spin in silicon}

\author{Cécile X. Yu}
\thanks{Contributed equally to the work.}
\affiliation{Univ. Grenoble Alpes, CEA, Grenoble INP, IRIG-Pheliqs, Grenoble, France.}
\author{Simon Zihlmann}
\thanks{Contributed equally to the work.}
\affiliation{Univ. Grenoble Alpes, CEA, Grenoble INP, IRIG-Pheliqs, Grenoble, France.}
\email{simon.zihlmann@cea.fr}
\author{José C. Abadillo-Uriel}
\affiliation{Univ. Grenoble Alpes, CEA, IRIG-MEM-L\_Sim, Grenoble, France.}
\author{Vincent P. Michal}
\affiliation{Univ. Grenoble Alpes, CEA, IRIG-MEM-L\_Sim, Grenoble, France.}
\author{Nils Rambal}
\affiliation{Univ. Grenoble Alpes, CEA, LETI, Minatec Campus, Grenoble, France.}
\author{Heimanu Niebojewski}
\affiliation{Univ. Grenoble Alpes, CEA, LETI, Minatec Campus, Grenoble, France.}
\author{Thomas Bedecarrats}
\affiliation{Univ. Grenoble Alpes, CEA, LETI, Minatec Campus, Grenoble, France.}
\author{Maud Vinet}
\affiliation{Univ. Grenoble Alpes, CEA, LETI, Minatec Campus, Grenoble, France.}
\author{Étienne Dumur}
\affiliation{Univ. Grenoble Alpes, CEA, Grenoble INP, IRIG-Pheliqs, Grenoble, France.}
\author{Michele Filippone}
\affiliation{Univ. Grenoble Alpes, CEA, IRIG-MEM-L\_Sim, Grenoble, France.}
\author{Benoit Bertrand}
\affiliation{Univ. Grenoble Alpes, CEA, LETI, Minatec Campus, Grenoble, France.}
\author{Silvano De Franceschi}
\affiliation{Univ. Grenoble Alpes, CEA, Grenoble INP, IRIG-Pheliqs, Grenoble, France.}
\author{Yann-Michel Niquet}
\affiliation{Univ. Grenoble Alpes, CEA, IRIG-MEM-L\_Sim, Grenoble, France.}
\author{Romain Maurand}
\affiliation{Univ. Grenoble Alpes, CEA, Grenoble INP, IRIG-Pheliqs, Grenoble, France.}
\email{romain.maurand@cea.fr}

\begin{abstract}
\textbf{Spins in semiconductor quantum dots constitute a promising platform for scalable quantum information processing. Coupling them strongly to the photonic modes of superconducting microwave resonators would enable fast non-demolition readout and long-range, on-chip connectivity, well beyond nearest-neighbor quantum interactions. Here we demonstrate strong coupling between a microwave photon in a superconducting resonator and a hole spin in a silicon-based double quantum dot issued from a foundry-compatible MOS fabrication process. By leveraging the strong spin-orbit interaction intrinsically present in the valence band of silicon, we achieve a spin-photon coupling rate as high as 330~MHz largely exceeding the combined spin-photon decoherence rate. This result, together with the recently demonstrated long coherence of hole spins in silicon, opens a new realistic pathway to the development of circuit quantum electrodynamics with spins in semiconductor quantum dots.}
\end{abstract}

\maketitle

Cavity quantum electrodynamics (QED) deals with the interaction between the quantum degrees of freedom of an atom and the electromagnetic modes of a  cavity~\cite{2006_Haroche}. The extension of this concept to superconducting quantum circuits has led to the development of circuit QED, opening new opportunities for the study of light-matter interaction and fostering  the progress of solid-state quantum processors based on superconducting qubits~\cite{2004_Wallraff,2004_Blais,2021_Blais}. In the same footsteps, a variety of alternative realizations have been explored using different types of quantum systems as artificial atoms~\cite{2020_Clerk}. Hybrid systems made of quantum dots coupled to superconducting microwave resonators are a prominent example~\cite{2004_Childress,Burkard2006,2012_Hu,Jin2012, 2012_Frey, 2012_Petersson, 2015_Viennot}. Of particular interest are silicon-based quantum dots owing to their ability to host long-coherence qubits encoded in a spin degree of freedom. Silicon-based spin qubits have made remarkable progress, reaching high fidelities in both one- and two-qubit gate operations, the latter being enabled by tunneling-mediated exchange interaction between neighboring qubits~\cite{2021_Burkard}. 
The co-integration with superconducting cavities acting as quantum buses would allow for long-range connectivity, largely facilitating the scalability of silicon spin qubits~\cite{1999_Imamoglu,2017_Vandersypen}.

As the spin does not directly couple to the cavity electric field, a spin-charge hybridization mechanism is needed to achieve coherent spin-photon interfaces. For electrons in Si/SiGe double quantum dots (DQDs), spin-photon coupling rates of a few tens of MHz have been demonstrated with the help of a synthetic spin-orbit (SO) interaction created by nearby micromagnets~\cite{2018_Samkharadze,2018_Mi,2019_Borjans,2022_HarveyCollard}. The reported coupling rates are several times larger than the spin dephasing rate, thereby enabling coherent spin-photon coupling ~\cite{2018_Samkharadze,2018_Mi} and cavity-mediated interaction between spins in distant DQDs ~\cite{2019_Borjans,2022_HarveyCollard}. Yet, to fully profit from circuit QED tools, including long-range, high-fidelity two-qubit operations and quantum non-demolition readout, a much stronger coupling strength is required calling for more efficient coupling schemes. 

In this work, we turn to a hole spin in a silicon-nanowire-MOS DQD in order to exploit the strong intrinsic SO interaction of valence band states~\cite{1955_Luttinger, 2003_Winkler}, whose potential for circuit QED~\cite{2013_Kloeffela,2017_Nigg,2021_Mutter,2022_Michal,2022_Bosco} has remained unexplored. In our device geometry, the quasi-one-dimensional hole confinement enhances this SO interaction~\cite{2018_Kloeffel} such that the SO length $\lso$, \textit{i.e.} the distance over which a spin rotates by $\pi$ due to SO interaction, is reduced to a few tens of nanometers, comparable to the DQD spatial extension $d$. The presence of such a strong SO interaction dramatically modifies the DQD energy levels resulting in the formation of a flopping-mode SO qubit~\cite{2021_Mutter} whose energy is well separated from the other excitations of the DQD system. Here, we demonstrate that this spin qubit interacts strongly with the quantized field of a high impedance superconducting microwave resonator. We observe a spin-photon coupling rate as large as $\gs/2\pi=\SI{330}{MHz}$, exceeding the combined spin and photon decoherence rate by a factor 27 and hence leading to a cooperativity of 1600. We furthermore explore the impact of the SO interaction in the limit of a spin confined in a single quantum dot. We measure a spin-photon coupling rate of $\sim \SI{1}{MHz}$ in line with recent predictions~\cite{2022_Bosco,2022_Michal}.

\section*{Hole spin circuit-QED architecture}
\begin{figure*}[htbp]
	\includegraphics[width=\textwidth]{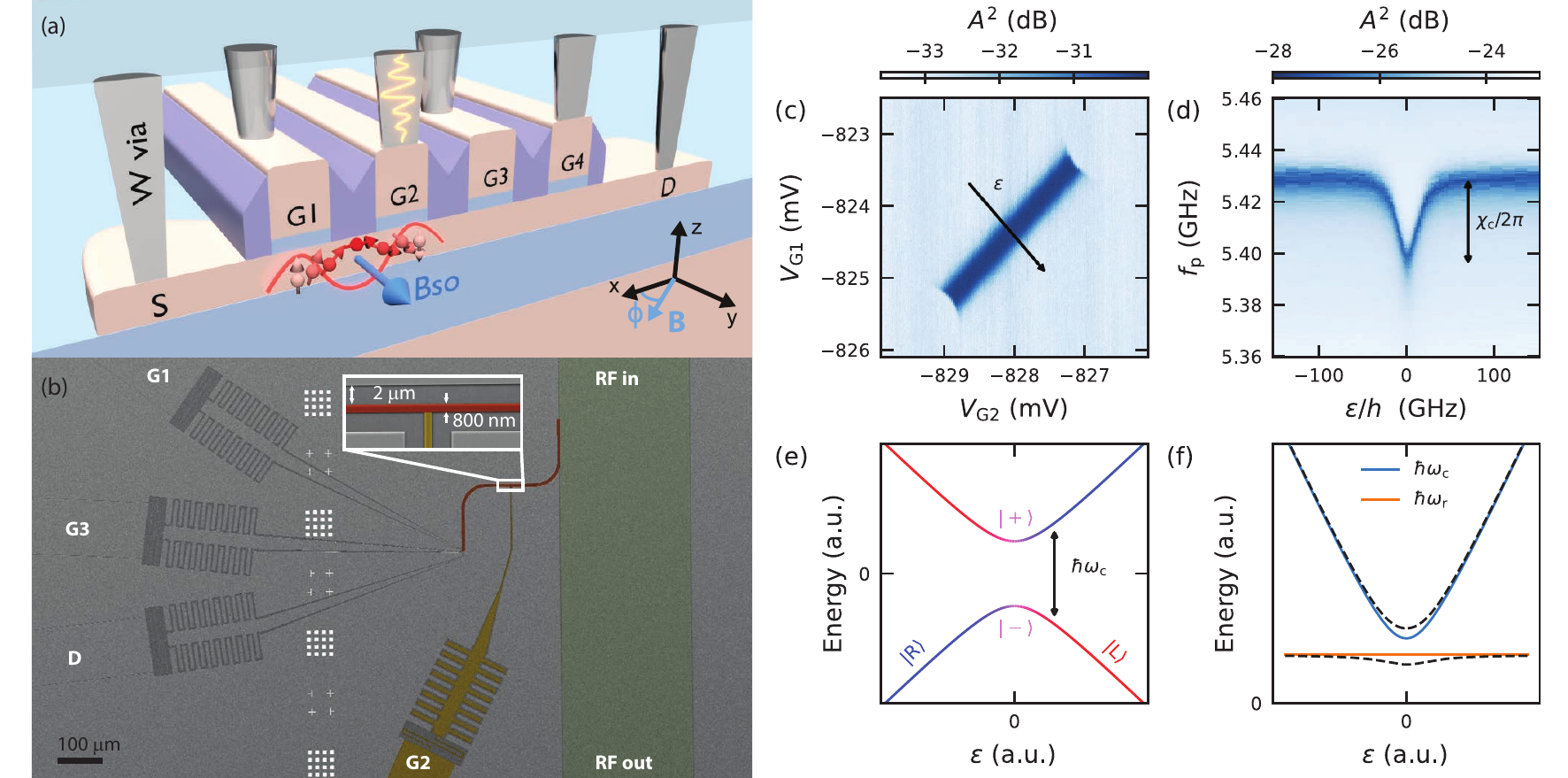}
	\caption{\label{fig:device}\textbf{Silicon MOS device, superconducting circuitry and DQD charge properties.}
	(a) Schematic cross section of the DQD device. The gates (G), the source (S) and the drain (D) of the nanowire transistor are connected with vertical vias to the NbN circuitry located at the surface of the device. We refer to the dot formed below G1 (G2) as the left (right) dot. The effect of the SO interaction is illustrated by a hole whose spin rotates coherently while tunneling from one dot to the other in the presence of an effective spin-orbit field $\mathbf{B}_\mathrm{so}$ (blue arrow). The orange wave-packet on the via connecting G2 pictures the photon in the microwave cavity. The external magnetic field $\mathbf{B}$ is applied in-plane with an angle $\phi$ to the nanowire axis. 
	(b) False-color top-view scanning electron micrograph of a representative NbN circuitry. The areas where the NbN is etched away appear in dark gray. The \SI{50}{\ohm} feed-line, the resonator and its DC bias line are highlighted in green, red and yellow, respectively. Inset: close-up of the DC tap at the center of the resonator. Each DC line is equipped with a LC low-pass filter consisting of an interdigitated capacitor and a nanowire inductor.
	(c) Transmission probed at the bare resonator resonance frequency (\SI{5.43}{GHz}) as a function of $\gone$ and $\gtwo$ at $\mathbf{B}=\mathbf{0}$. The energy detuning $\varepsilon$ between the two dots is swept along the black arrow.
	(d) Transmission as a function of $f_\mathrm{p}$ and $\varepsilon$. At large $|\varepsilon|$ the bare resonator is probed, whereas near $\varepsilon=0$, the DQD charge qubit interacts dispersively with the resonator leading to a frequency shift $\chi_\mathrm{c}/2\pi$.
	(e) Qualitative energy diagram of the DQD as a function of the energy detuning $\varepsilon$. The charge can be either in the right dot $\ket{\mathrm{R}}$ (blue) or in the left dot $\ket{\mathrm{L}}$ (red). Due to the finite tunnel coupling $t_\mathrm{c}$ between the dots, $\ket{\mathrm{L}}$ and $\ket{\mathrm{R}}$ hybridize near $\varepsilon=0$ into a bonding $\ket{-}$ and an antibonding $\ket{+}$ state, forming the basis of a charge qubit with energy $\hbar\omega_\mathrm{c}$.
	(f) Qualitative energy diagram of the charge qubit energy and the cavity energy as a function of $\varepsilon$. The colored solid lines correspond to the non-interacting case, while the dashed lines sketch the dispersive repulsion experienced by the cavity and the charge qubit in the presence of a finite charge-photon coupling. } 
\end{figure*}

The spin circuit QED architecture we have designed features a hole confined in a silicon DQD device interacting with a single microwave photon trapped in a superconducting cavity. The DQD is hosted in a natural silicon nanowire MOS transistor whose channel is controlled by four $\Omega$-shape gates crossing the  nanowire, as shown in Fig.~\ref{fig:device}~(a). The back-end-of-line fabrication of the silicon chip is interrupted in order to replace the first metallic interconnect layer with a 10-nm-thick NbN layer with large kinetic inductance \cite{2019_Niepce, 2021_Yu} and magnetic field resilience \cite{2021_Yu}. A high-impedance (\SI{2.5}{k\ohm}) microwave cavity is then patterned in the NbN film, along with a \SI{50}{\ohm} microwave feed-line, ground planes and fanout lines (see Fig.~\ref{fig:device}~(b)). Besides being well suited for future large-scale integration, the used foundry-compatible MOS technology comes with large gate capacitances resulting in a tight electrostatic control and hence strong coupling to the electric-field component of the cavity mode.

Negative voltages applied to G1 and G2 accumulate holes in a DQD potential, as illustrated in Fig.~\ref{fig:device}~(a).
The so-confined holes in the valence band of silicon experience a strong SO interaction~\cite{1955_Luttinger, 2003_Winkler}, which, along with an applied in-plane magnetic field $\mathbf{B}$, controls the spin-charge mixing in the DQD, as discussed hereafter.

We probe the microwave response of this hybrid system in transmission at a temperature of \SI{8}{mK} and at powers corresponding to less than one photon on average in the cavity ($n_{\mathrm{avg}}\sim0.1$), which is assumed to be in its ground state. We first characterize the bare cavity response by sweeping the probe frequency $f_\mathrm{p}=\omega_\mathrm{p}/2\pi$ across the resonance frequency while keeping the charges in the DQD fixed. That way we extract a bare cavity resonance frequency $f_\mathrm{r}=\omega_\mathrm{r}/2\pi=\SI{5.428}{GHz}$ and a cavity decay rate $\kappa/2\pi=\SI{14}{MHz}$. To characterize the charge-photon coupling strength $\gc$, we monitor the transmission at frequency $f_\mathrm{r}$ while $\gone$ and $\gtwo$ are varied, see Fig.~\ref{fig:device}~(c). In the dark blue region, the levels of the two dots are aligned so that a hole oscillates between the dots in response to the cavity electric field. We next probe the transmission as a function of $f_\mathrm{p}$ and the energy detuning $\varepsilon$ between the two dots, see Fig.~\ref{fig:device}~(d). This reveals a dispersive downshift of the cavity resonance near $\varepsilon=0$, due to the electric dipole interaction with the DQD hole charge qubit \cite{2004_Blais, 2005_Gorman} with energy $\hbar\omega_\mathrm{c}=\sqrt{\varepsilon^2 + 4 t_\mathrm{c}^2}>\hbar\omega_\mathrm{r}$ (see Figs.~\ref{fig:device}~(e) and (f)). From the temperature dependence of this dispersive shift, we extract a charge-photon coupling strength $\gc/2\pi=\SI{513}{MHz}$ together with an interdot tunnel coupling $t_\mathrm{c}/h=\SI{9.57}{\giga \hertz}$ (see Supplementary Information section IV).

\section*{Strong hole spin-photon coupling}
\begin{figure}[ht]
	\includegraphics[scale=1]{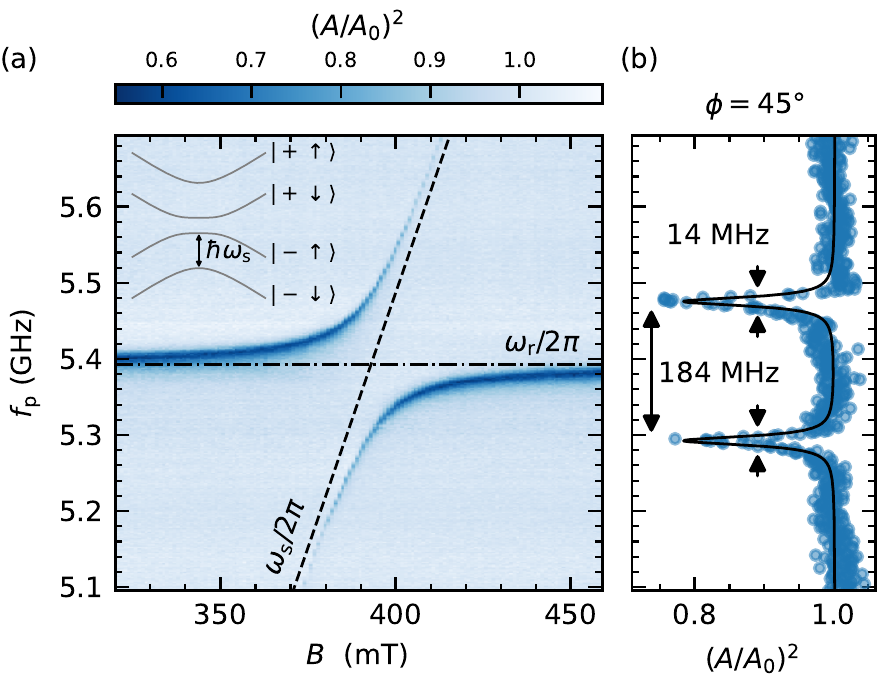}
	\caption{\label{fig:anticrossing}\textbf{Strong spin-photon coupling.} (a) Normalized transmission as a function of the probe frequency $f_\mathrm{p}$ and the amplitude of the magnetic field $\mathbf{B}$ oriented at $\phi=\SI{45}{\degree}$ with respect to the nanowire axis. Inset: qualitative energy diagram of the DQD at finite magnetic field. The magnetic field splits the spin degenerate $\ket{-}$ and $\ket{+}$ states (see Fig.~\ref{fig:device}~(e)) into four states $\ket{-\downarrow}$, $\ket{-\uparrow}$, $\ket{+\downarrow}$ and $\ket{+\uparrow}$. An avoided crossing, signature of strong spin-photon coupling, is observed when the spin transition frequency $\omega_\mathrm{s}/2\pi$ (dashed line) matches the bare resonator frequency $\omega_\mathrm{r}/2\pi$ (dashed-dotted line). (b) Frequency line cut at resonance highlighting a vacuum Rabi mode splitting $2\gs/2\pi=\SI{184}{MHz}$. The solid line is a fit to two superimposed Lorentzian functions, whose width (\SI{14}{M\hertz}) is twice the decoherence rate of the hybridized spin-photon state.
	}
\end{figure}

An in-plane magnetic field $\mathbf{B}$ lifts the spin degeneracy of the DQD charge states as sketched in the inset of Fig.~\ref{fig:anticrossing} (a). The  two lowest spin-polarized states define a spin-orbit flopping mode qubit~\cite{2021_Mutter}. We probe the spin-photon interaction at $\varepsilon\sim 0$, where the electric dipole of the hole in the DQD is maximal.
When the Zeeman spin splitting $\hbar\omega_\mathrm{s}$ matches the resonance frequency of the cavity ($f_\mathrm{r}\simeq f_\mathrm{s}=\omega_\mathrm{s}/2\pi$), spin-photon hybridization results in an avoided crossing which splits the cavity response into two branches separated by the vacuum Rabi mode splitting~\cite{2004_Wallraff, 2004_Blais}.
A representative measurement of this avoided crossing is shown in Fig.~\ref{fig:anticrossing} (a), where the normalized transmission is plotted as a function of $f_\mathrm{p}$ and $B=|\mathbf{B}|$ at $\phi=\SI{45}{\degree}$ with respect to the nanowire axis. The line-cut at resonance shows two distinct dips separated by a vacuum Rabi mode splitting $2\gs/2\pi=\SI{184}{MHz}$, with $\gs$ the spin-photon coupling. The line-width of these dips yields the decoherence rate of the hybridized spin-photon states $\frac{1}{2}(\gamma_\mathrm{s}+\kappa/2)/2\pi=\SI{7}{MHz}$ with $\gamma_\mathrm{s}$ the spin decoherence rate. From $\kappa/2\pi=\SI{14}{M\hertz}$, we extract $\gamma_\mathrm{s}/2\pi=\SI{7}{MHz}$. The fact that $\gs\gg\kappa,\gamma_\mathrm{s}$ demonstrates a strong coupling between the hole spin and the photon in the cavity. 

\section*{Spin-photon coupling vs magnetic-field orientation}
\begin{figure*}[htbp]
	\includegraphics[width=\textwidth]{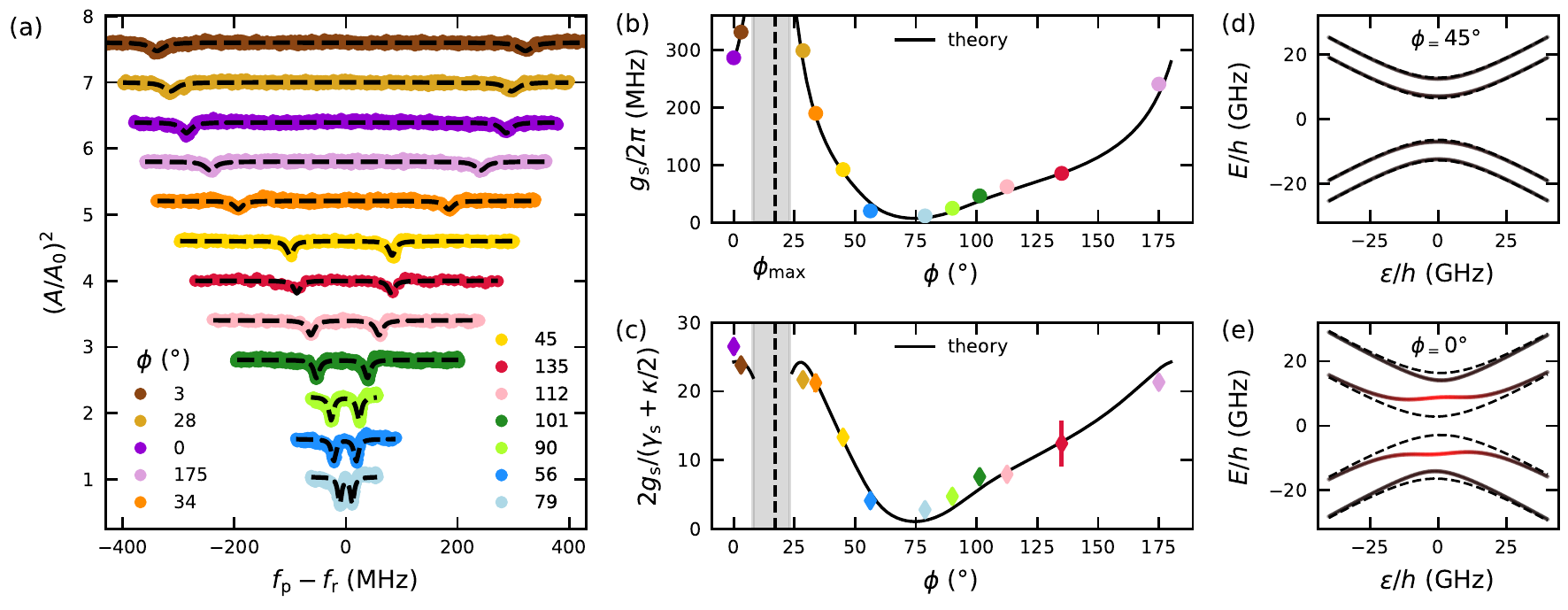}
	\caption{\label{fig:angular_dpendence}\textbf{Spin-photon coupling vs magnetic-field orientation.} (a) Normalized transmission as a function of the probe frequency $f_\mathrm{p}$ for various magnetic field orientations $\phi$. All curves are measured at resonance (i.e. for $f_\mathrm{s}=f_\mathrm{r}$) and all show a clear vacuum Rabi mode splitting. The curves are offset vertically and centered around the cavity resonance frequency $f_\mathrm{r}$ for clarity. The dashed lines are fits to a superposition of two Lorentzians. For each vacuum Rabi mode cut shown here, a full map of the avoided crossing similar to Fig.~\ref{fig:anticrossing} is shown in the Supplementary Figure S8.
	(b) Angular dependence of the spin-photon coupling $\gs$ and (c) $2\gs/(\gamma_\mathrm{s}+\kappa/2)$. The experimental data are in excellent agreement with the theory (solid line). The grey shaded area outlines the magnetic field orientations where the spin-photon resonance is achieved for magnetic fields larger than \SI{1}{T}, which are inaccessible in our experimental setup. The vertical dashed line indicates the magnetic field angle $\phi_\mathrm{max}=\SI{17}{\degree}$ at which the spin-photon coupling is maximal. The error bars in (b) and (c) represent the standard deviation from the fitting.
	Energy diagram of the DQD at resonance for $\phi=\SI{45}{\degree}$ (d) and $\phi=\SI{0}{\degree}$ (e). The dashed lines represent the spin states in the absence of SO interaction (see also inset to Fig.~\ref{fig:anticrossing}~(a)), where the spin-splitting energies are, in a first approximation, independent of the DQD detuning. The SO interaction primarily couples the dashed $\ket{-\uparrow}$ and $\ket{+\downarrow}$ states, as highlighted by the color of the solid lines (increasing mixing from black to red). Spin-charge mixing leads to pronounced detuning dependence of the spin-splitting energies. Notice that for $\phi=\SI{0}{\degree}$ (panel (e)) the spin-charge mixing is stronger than for $\phi=\SI{45}{\degree}$ (panel (d)), which results in a larger spin-photon coupling $\gs$ (see panel (b)).
	}
\end{figure*}

Varying the orientation of the in-plane magnetic field reveals a pronounced anisotropy in the vacuum Rabi mode splitting with a measured maximal (minimal) $\gs/2\pi$ of \SI{330}{MHz} (\SI{10}{MHz}) at an angle $\phi=\SI{3}{\degree}$ ($\phi=\SI{79}{\degree}$), as shown in Fig.~\ref{fig:angular_dpendence}~(a) and (b). This large modulation in the spin-photon coupling strength results from the interplay between the Zeeman effect and the SO interaction, leading to $\gs \propto \gc |({\mathsf{g}}\mathbf{B})\cross(\mathsf{g}\mathbf{B}_\mathrm{so})|$ with $\mathsf{g}$ the average gyromagnetic $\mathsf{g}$-matrix of the two dots and $\mathbf{B}_\mathrm{so}$ the effective spin-orbit field (note that $|\mathbf{B}_\mathrm{so}|$ is related to the spin-orbit length, see Supplementary Information section IX). The spin-photon coupling is thus expected to vanish when $\mathbf{B}$ is parallel to $\mathbf{B}_\mathrm{so}$, and to be maximum at an angle $\phi_\mathrm{max}$ where the spin Larmor vector $\mathsf{g}\mathbf{B}$ is approximately perpendicular to $\mathsf{g}\mathbf{B}_\mathrm{so}$. The $\mathsf{g}$-matrix of holes embodies an anisotropic Zeeman splitting $E_\mathrm{Z}=\mu_\mathrm{B}|\mathsf{g}\mathbf{B}|$, in contrast to electrons ($E_\mathrm{Z}\approx 2\mu_\mathrm{B}B$)~\cite{2018_Crippa}. In the present case, both dots show similar $\mathsf{g}$-matrix anisotropies, with $E_\mathrm{Z}\simeq1.3\mu_\mathrm{B}B$ when $\mathbf{B}$ is along the $x$-axis, and $E_\mathrm{Z}\simeq2\mu_\mathrm{B}B$ when $\mathbf{B}$ is along the $y$-axis. As shown in Fig.~\ref{fig:angular_dpendence}~(b), $\gs/2\pi$ gets almost entirely suppressed around $\phi=\SI{75}{\degree}$. Since $\gs$ does not vanish completely, $\mathbf{B}_\mathrm{so}$ must have a small out-of-plane component. Overall, however, we can conclude that the orientation of $\mathbf{B}_\mathrm{so}$ is rather close to the $y$-axis. As discussed in the Supplementary Information section IX, the orientation of $\mathbf{B}_\mathrm{so}$ is primarily determined by the device geometry. It is expected to be perpendicular to the inter-dot tunneling direction ($\sim x$ axis) as well as to the average electric-field direction in the inter-dot barrier region ($\sim z$ axis). Due to the anisotropy of the $\mathsf{g}$-matrix, the magnetic field orientation maximizing $\gs$ is not exactly orthogonal to $\mathbf{B}_\mathrm{so}$.

The quality of the spin-photon interface can be further quantified by the ratio between the coupling strength and the decoherence rate, i.e. $2\gs/(\gamma_\mathrm{s} + \kappa/2)$, which we plot in Fig.~\ref{fig:angular_dpendence}~(c). This ratio reaches up to 27 for $\phi=\SI{0}{\degree}$, which, along with a cooperativity \cite{2020_Clerk} $C=4\gs^2/(\gamma_\mathrm{s}\kappa)=1600$, highlights an extremely strong light-matter interaction.
To evaluate the relative impact of spin dephasing and cavity decay rates, we can extract the angular dependence of $\gamma_\mathrm{s}/2\pi$ (see Supplementary Information section VIII). We find values ranging from \SI{2.5}{MHz} to \SI{17}{MHz}. Small (large) spin dephasing generally coincides with a small (large) spin-charge mixing. Over a large angular range centered around $\phi=\SI{75}{\degree}$, $\gamma_\mathrm{s}$ remains rather small and the quality of the spin-photon interface is mostly limited by the relatively large cavity decay rate $\kappa/2\pi= \SI{14}{MHz}$.

A model for the spin-photon coupling of a hole DQD with SO interaction is presented in the Supplementary Information section IX and X. With distinct anisotropic Zeeman response for the two quantum dots and spin-dependent tunnel couplings, this model captures $\gs$ at all magnetic field orientations as shown in Fig.~\ref{fig:angular_dpendence}~(b). The effect of spin charge-mixing on the DQD energy diagram is illustrated  in Figs.~\ref{fig:angular_dpendence}~(d) and (e) for two different magnetic field orientations. In particular, Fig.~\ref{fig:angular_dpendence}~(e) highlights the large renormalization of the energy levels in a DQD due to strong SO interaction. Our model also catches the decrease of coherence by spin-charge mixing as the magnetic field orientation approaches $\phi_{\mathrm{max}}$, as shown in Fig.~\ref{fig:angular_dpendence}~(c) and in the Supplementary Information section XI. The black solid line is calculated assuming a charge qubit decoherence rate $\gamma_{\mathrm{c}}/2\pi=\SI{9.9}{M\hertz}$, due to the dominant detuning noise, on top of an isotropic bare spin decoherence rate  $\gamma_0/2\pi=\SI{3.4}{M\hertz}$, collecting other mechanisms such as nuclear-spin noise~\cite{fischer2008spin}, electrically induced $\gt$-factor fluctuations \cite{2022_Piot, 2022_Michal}, or phonon-mediated spin relaxation~\cite{li2020hole}.

\section*{Spin-photon coupling in the single-dot limit}
\begin{figure}[htbp]
	\includegraphics[scale=1]{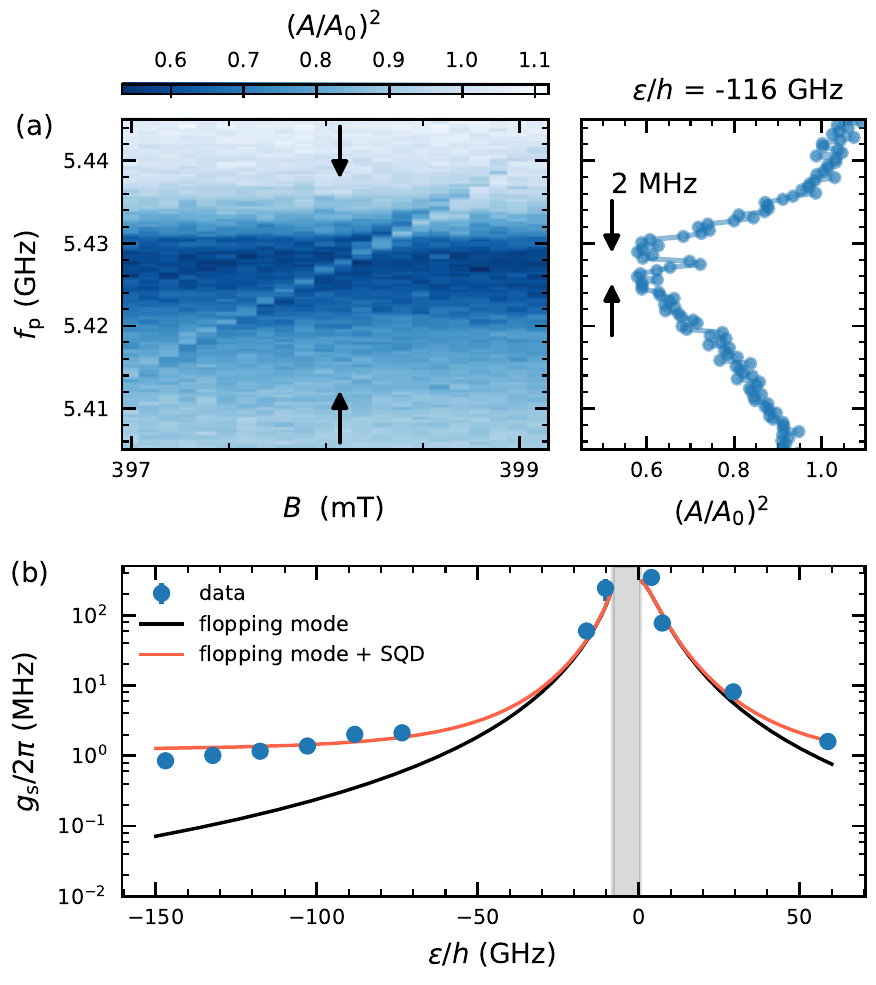}
	\caption{\label{fig:single_dot}\textbf{Spin-photon coupling in the single-dot limit.} (a) Left panel: normalized transmission as a function of $f_\mathrm{p}$ and $B$ at $\varepsilon=\SI{-116}{GHz}$ and $\phi=\SI{11.25}{\degree}$. Right panel: frequency line cut at the position of the arrows in the left panel. The spin transition is clearly visible and a double-dip structure split by $\sim\SI{2}{MHz}$ is observed at resonance. (b) $\gs/2\pi$ as a function of energy detuning $\varepsilon$. The DQD flopping mode model (black solid line) captures the large spin-photon coupling around $\varepsilon \sim 0$ very well, but underestimates $\gs$ by more than one order of magnitude at large $\varepsilon$, where the hole is confined in a single dot (SQD) on the left ($\varepsilon>0$) or on the right ($\varepsilon<0$). The measured $\gs$ is very well reproduced by adding an asymptotic single-dot  spin-photon coupling $\gs^{(R, L)}$ in each QD. We find $\gs^{(R)}=\SI{1.16}{MHz}$ and $\gs^{(L)}=\SI{0.66}{MHz}$, as shown by the red solid line. The grey shaded area outlines the energy detuning range where the spin-photon resonance is achieved for magnetic fields larger than \SI{1}{T}, which are inaccessible in our experimental setup. The error bars in represent the standard deviation from the fitting.}
\end{figure}

The spin-photon couplings reported so far benefit from the large electric dipole moment of the DQD at $\varepsilon\sim 0$. In the following, we explore the interaction between the spin and the cavity when the hole is localized in a single quantum dot. As previously reported for electrons~\cite{2018_Mi}, this interaction quickly vanishes for increasing detuning since charge localization in a single dot quenches the electric dipole moment. This is evidenced by the detuning dependence of the charge-photon coupling  $\gc^{\mathrm{eff}}=2\gc t_\mathrm{c}/\sqrt{4t_\mathrm{c}^2+\varepsilon^2}$. Fig.~\ref{fig:single_dot}~(a) shows the transmission when the microwave cavity is resonant with the spin splitting of the hole confined in the right dot ($\varepsilon \ll -t_\mathrm{c}$). Despite the considerable reduction of the hole dipole moment, two dips are still visible in the transmission, implying $\gamma_{\mathrm{s}} < \gs$. Their separation reveals a spin-photon coupling $\gs/2\pi\sim\SI{1}{\mega\hertz}$. Since  $\gs < \kappa$ (the so-called bad cavity limit), a clear vacuum Rabi mode splitting cannot be resolved. To further support the existence of a single-dot spin-photon interaction, we measure $\gs$ as a function of $\varepsilon$, see Fig.~\ref{fig:single_dot}~(b). We find that $\gs$ drops by more than two orders of magnitude when increasing $|\varepsilon|$, but tends to saturate once the hole is fully localized in the right dot. This limit presents potential interest since the use of single dots \cite{2022_Michal,2022_Bosco} would simplify the device architecture, reduce the number of control parameters, allow for longer hole-spin coherence, and enable alternative and possibly more efficient spin-photon architectures~\cite{2017_Vandersypen,2017_Nigg}. In addition, significant progress can be expected from the implementation of spin-photon coupling schemes relying on operational sweets spots~\cite{2013_Kloeffela,2017_Nigg,2022_Michal,2022_Bosco}, where decoherence is reduced while preserving efficient electrical control.

\section*{Impact of strong spin-orbit interaction}
The large spin-photon cooperativity observed in our experiment can be seen as a combined effect of an efficient charge-photon coupling, favored by the MOS device layout, and the intrinsic SO interaction of holes, much stronger than the synthetic SO interaction of electrons in silicon.  We note that the maximum $\gs$ is generally achieved when $2t_\mathrm{c} \sim \hbar\omega_\mathrm{r} \sim E_\mathrm{z}$. Under this condition, the excited states $\ket{-\uparrow}$ and $\ket{+\downarrow}$ completely mix and $\gs$ approaches $\gc$, regardless of the SO interaction strength. In the case of electrons in silicon, however, the weak SO interaction ($\lso\gg d$) cannot keep these two states sufficiently apart from each other to prevent unwanted excitations to the second excited state $\ket{+\downarrow}$~\cite{2017_Benito}. This makes the limit of strong spin-charge mixing impractical thereby preventing the achievement of maximal spin-photon coupling. The hole system studied here does not suffer from this limitation. The level repulsion induced by the stronger SO interaction ($\lso\sim d$) keeps the $\ket{-\uparrow}$ and $\ket{+\downarrow}$ states apart, as illustrated in Fig. \ref{fig:angular_dpendence}~(e). 
Noticeably, the zero-detuning excitation energy of the flopping mode qubit gets significantly reduced as compared to Zeeman splitting in the absence of SO interaction. This implies that the spin-photon resonance condition occurs for $2 t_\mathrm{c} > \hbar \omega_\mathrm{r}$. In this regime, the flopping mode qubit is well isolated from the higher energy levels and strongly coupled to the cavity mode. Moreover, operating the DQD at larger tunnel coupling  ($2 t_\mathrm{c} \sim 4 \hbar \omega_\mathrm{r} $ in Fig.~\ref{fig:angular_dpendence}~(e)) reduces the impact of charge noise on the detuning energy thereby enabling large cooperativity.

\section*{Conclusions}
Looking further ahead, we foresee ample room to improve the spin-photon interface. On an engineering level, largely reduced resonator losses with $\kappa/2\pi<$~\SI{1}{MHz} should be readily feasible~\cite{2017_Mia,2020_Harvey-Collard,2021_Yu}, and further improvements harnessing the advanced MOS fabrication platform to integrate a multi-layer  superconducting back-end-of-line would allow for a well-controlled  microwave environment. We should also like to emphasize that, in our hole system, the spin-photon coupling strength is ruled by the geometry and electrostatic design of the DQD (see Supplementary Information section IX), and controlled by the amplitude and direction of the externally applied magnetic field. We expect this should limit the impact of device-to-device variability and facilitate the development of large-scale quantum networks.

Our work promotes holes in Si-MOS devices as a powerful playground for the development of spin circuit QED. As opposed to electrons, holes benefit from an intrinsically strong and versatile SO interaction ($\lso\sim d$), which is available without the need of micromagnets. This not only simplifies the device architecture, but also allows to engineer a SO two-level system, which remains well isolated from other energy levels of the DQD. Our demonstration of such a hole flopping mode spin qubit, with a unprecedented spin-photon cooperativity of 1600, opens the door to a wide range of circuit QED implementations of fundamental and practical interest (e.g., two-qubit operations between distant hole spins with fidelities as high as 90\% seem already within reach~\cite{2019_Benito}).

\clearpage

\section*{Acknowledgments}
We thank J.-L. Thomassin and F. Gustavo for help in the fabrication of the NbN circuitry and M. Boujard and I. Matei for technical support in the lab. Vincent Renard is acknowledged for careful proofreading of the manuscript.

This research has been supported by the European Union’s Horizon 2020 research and innovation programme under grant agreements No. 951852 (QLSI project), No. 810504 (ERC project QuCube) and  No. 759388 (ERC project LONGSPIN), and by the French National Research Agency (ANR) through the project MAQSi. S. Zihlmann acknowledges support by an Early Postdoc Mobility fellowship (P2BSP2\_184387) from the Swiss National Science Foundation.

\section*{Author contributions statement:}
C.Y. fabricated the NbN-circuitry with the help from S.Z.. C.Y. and S.Z. performed the measurements. S.Z. analyzed the data with inputs from C.Y., J.C.A.U., E.D. and R.M.. J.C.A.U. developed the theoretical model with the help from V.P.M., M.F., and Y.M.N.. S.Z., R.M., J.C.A.U., S.D.F. and Y.M.N. co-wrote the manuscript with inputs from all authors. N.R., H.N, T.B., M.V. and B.B. were responsible for the front-end fabrication of the device. R.M. initiated the project.

\section*{Competing Interests Statement}
M.V. is co-founder and CEO of siquance.

\section*{Data availability statement}
The datasets generated during the current study, including the code to analyze them, are available at \url{https://doi.org/10.5281/zenodo.7533669}.

\section*{Corresponding author email address:}
simon.zihlmann@cea.fr, romain.maurand@cea.fr

\clearpage

\setcounter{section}{0}
\setcounter{equation}{0}
\setcounter{figure}{0}
\setcounter{table}{0}

\renewcommand\thesection{S\arabic{section}}
\def\theequation{S\arabic{equation}}
\renewcommand{\thetable}{S\Roman{table}}
\renewcommand{\thefigure}{S\arabic{figure}}
\renewcommand{\theHfigure}{S\arabic{figure}}

\begin{center}
\textbf{\large Supplementary information for ``Strong coupling between a photon and a hole spin in silicon''}
\end{center}
\section{\label{sec:device}Device fabrication and measurements}
\begin{figure*}[htbp]
	\includegraphics[width=\textwidth]{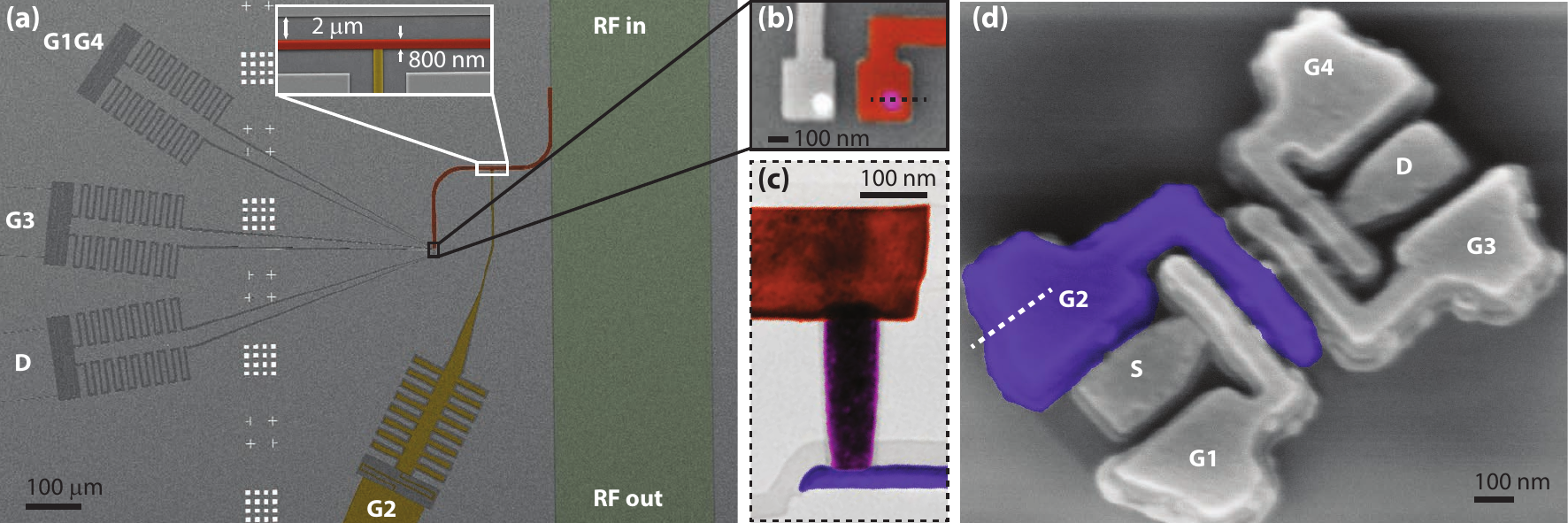}
	\caption{\label{fig:device_supp}\textbf{Device.} False color scanning (SEM) and transmission (TEM) electron micrographs of various parts of the device. (a) Overview with the feed line in green, the resonator in red and its DC-voltage tap in yellow. The inset is a zoom-in on the voltage tap that shows the dimension of the resonator (\SI{800}{nm} wide central conductor with \SI{2}{\micro m} gap). (b) Zoom-in on the connection of a NbN line to a vertical tungsten via (purple) emerging at the surface. (c) TEM image of a similar device, showing a representative tungsten via (purple) connecting the transistor (blue) to a classical, \SI{150}{nm} thick copper metallisation at M1 (red). In the actual device, the copper layer is replaced by a \SI{10}{nm} thick NbN layer, see (b). (d) Zoom-in on the device just after transistor fabrication. The dashed lines in (b) and (d) indicate the cuts corresponding to the TEM image in (c).}
\end{figure*}

The double quantum dot (DQD) is hosted in a silicon on insulator nanowire transistor with four $\Omega$-shaped gates in series. A false-color scanning electron micrograph of a nominally identical device is shown in Fig.~\ref{fig:device_supp}~(d). The natural Si channel is \SI{10}{nm} thick and \SI{40}{nm} wide. Each gate is \SI{40}{nm} long and the inter-gate separation is \SI{40}{nm} as well. The gates are insulated from the channel by a \SI{5}{nm} thick layer of thermal SiO$_2$. Boron doped source and drain are used as hole reservoirs. The industrial fabrication on \SI{300}{mm} wafers~\cite{2012_Barraud} is interrupted just before the first metallic interconnect layer (M1). The wafer is then flattened by chemical-mechanical polishing (CMP). Tungsten vias connecting to source, drain and the gates $\sim\SI{200}{nm}$ below the pre-metal dielectric (PMD) are exposed at the wafer surface, as shown in Figs.~\ref{fig:device_supp}~(b,~c). At this stage, the resonator as well as all DC connections are fabricated by sputter deposition of a \SI{10}{nm} thick NbN layer with subsequent patterning and etching in a SF$_6$/O$_2$ plasma in an academic clean room, see Fig.~\ref{fig:device_supp}~(a). Further details on the NbN resonator characteristics are provided elsewhere \cite{2021_Yu}. All DC connections are fitted with LC low pass filters consisting of an interdigitated capacitor of \SI{0.134}{pF} and a nanowire inductor of \SI{123}{nH} resulting in a cut-off frequency of \SI{1.2}{GHz}, similar to previously reported filters \cite{2017_Mia,2020_Harvey-Collard}. In order to reduce the number of DC lines fanning out from the device, the source is hard grounded to the NbN ground plane and G1 and G4 are shorted at the device level (NbN circuitry). G2 is galvanically connected to the voltage anti-node of the resonator. Before measurements, the chip is annealed in forming gas (N$_2$/H$_2$ 4\%) at \SI{400}{\degreeCelsius} for 1 hour. 

\begin{figure*}[htbp]
	\includegraphics[width=10cm]{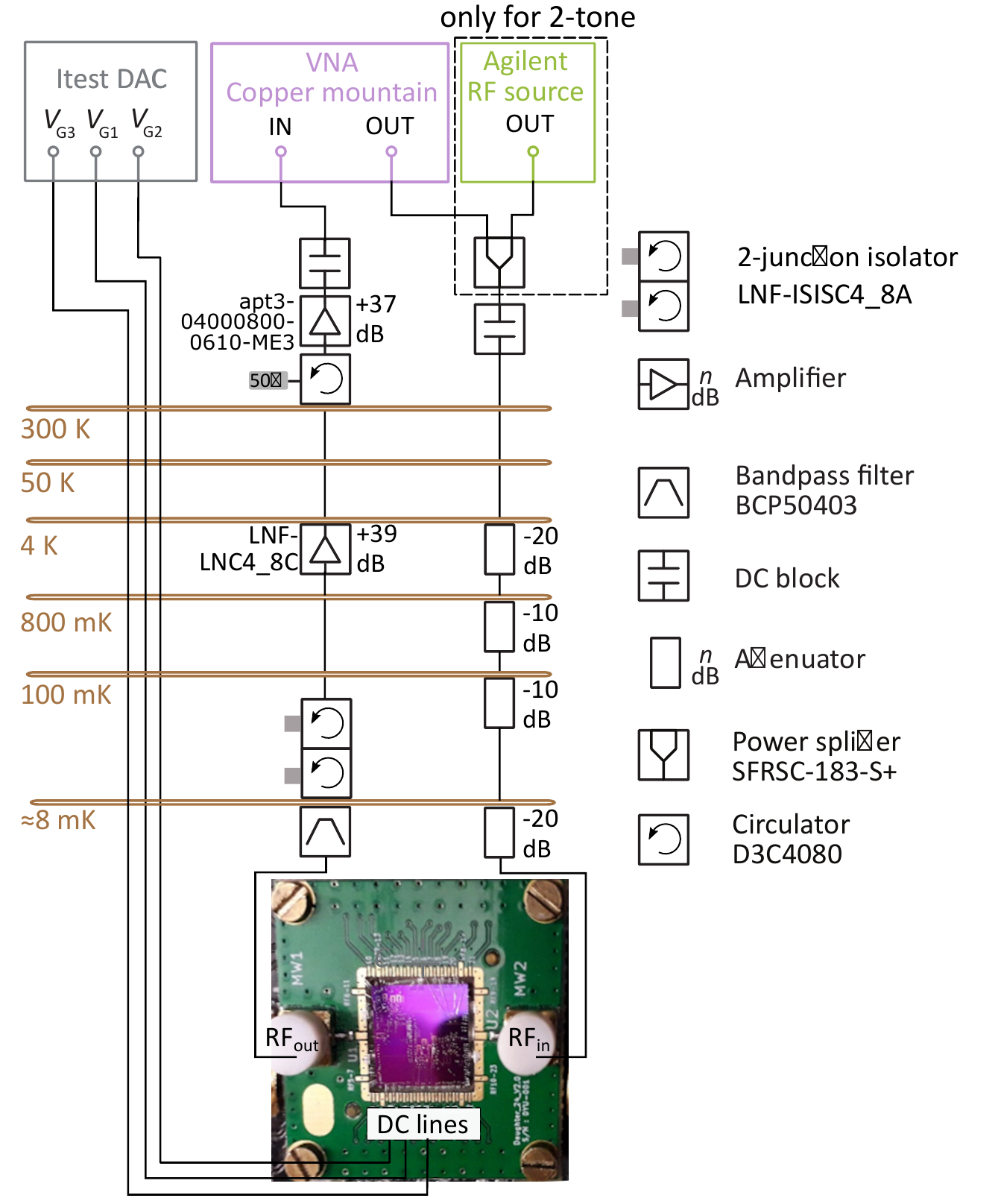}
	\caption{\label{fig:setup}\textbf{Measurement setup.}}
\end{figure*}

All measurements are performed in a dilution refrigerator equipped with a three axes vector magnet, at a base temperature $T=\SI{8}{mK}$. The input line of the resonator has \SI{60}{dB} discrete attenuation and the read-out line is equipped with a standard cryogenic low noise amplifier and a second amplifier at room temperature, see Fig.~\ref{fig:setup}. The transmission $A^2$ of the system is inferred from the complex scattering parameter $A=|S_{21}|$ measured with a VNA from Copper Mountain, model M5180. The DC gate voltages are supplied by a BE2231 card in a Bilt rack from Itest and are low pass filtered at mixing chamber temperature (multi stage LC and RC filters). For two-tone spectroscopy measurements, the second tone is generated by an Agilent E8527D, whose clock is synchronized with the VNA clock. In order to reach the single photon limit in the microwave cavity (\SI{\sim-120}{dBm}), an other \SI{-20}{dB} attenuator is added at the output port of the VNA. 

\section{\label{sec:cavity}Cavity characterization}
\begin{figure*}[htbp]
	\includegraphics[scale=1]{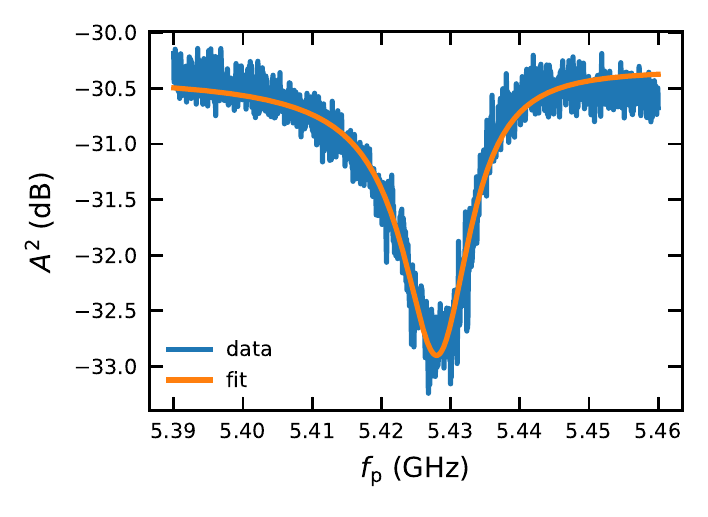}
	\caption{\label{fig:cavity}\textbf{Cavity characterization.} Transmission through the feedline as a function of probe frequency with less then one photon on average in the resonator at zero magnetic field and with the charge qubit largely detuned from the resonator.}
\end{figure*}

The transmission ($A^2$) is plotted in Fig.~\ref{fig:cavity} as a function of the probe frequency $f_\mathrm{p}$ close to the resonator frequency. The probe power applied at the resonator input is \SI{-130}{dBm}, which corresponds to well below one photon on average in the cavity ($n_\mathrm{avg}\sim0.1$). Therefore, the cavity can be assumed to be in its ground state. All following experiments were conducted at such low microwave powers. The magnetic field is set to zero, and the gates are biased so that no hole is able to move between the two dots ($\gone=\SI{-0.822}{V}$ and $\gtwo=\SI{-0.826}{V}$). By fitting the transmission \cite{2012_Megrant}, we extract the resonance frequency $\omega_\mathrm{r}/2\pi=\SI{5.428}{GHz}$, as well as the internal and external quality factors $Q_\mathrm{int}=530$ and $Q_\mathrm{ext}=1550$, which correspond to cavity decay rates $\kappa_\mathrm{int}/2\pi=\SI{10}{MHz}$ and $\kappa_\mathrm{ext}/2\pi=\SI{3.5}{MHz}$.

\section{\label{sec:stability}Charge stability diagram}
\begin{figure*}[htbp]
	\includegraphics[width=\textwidth]{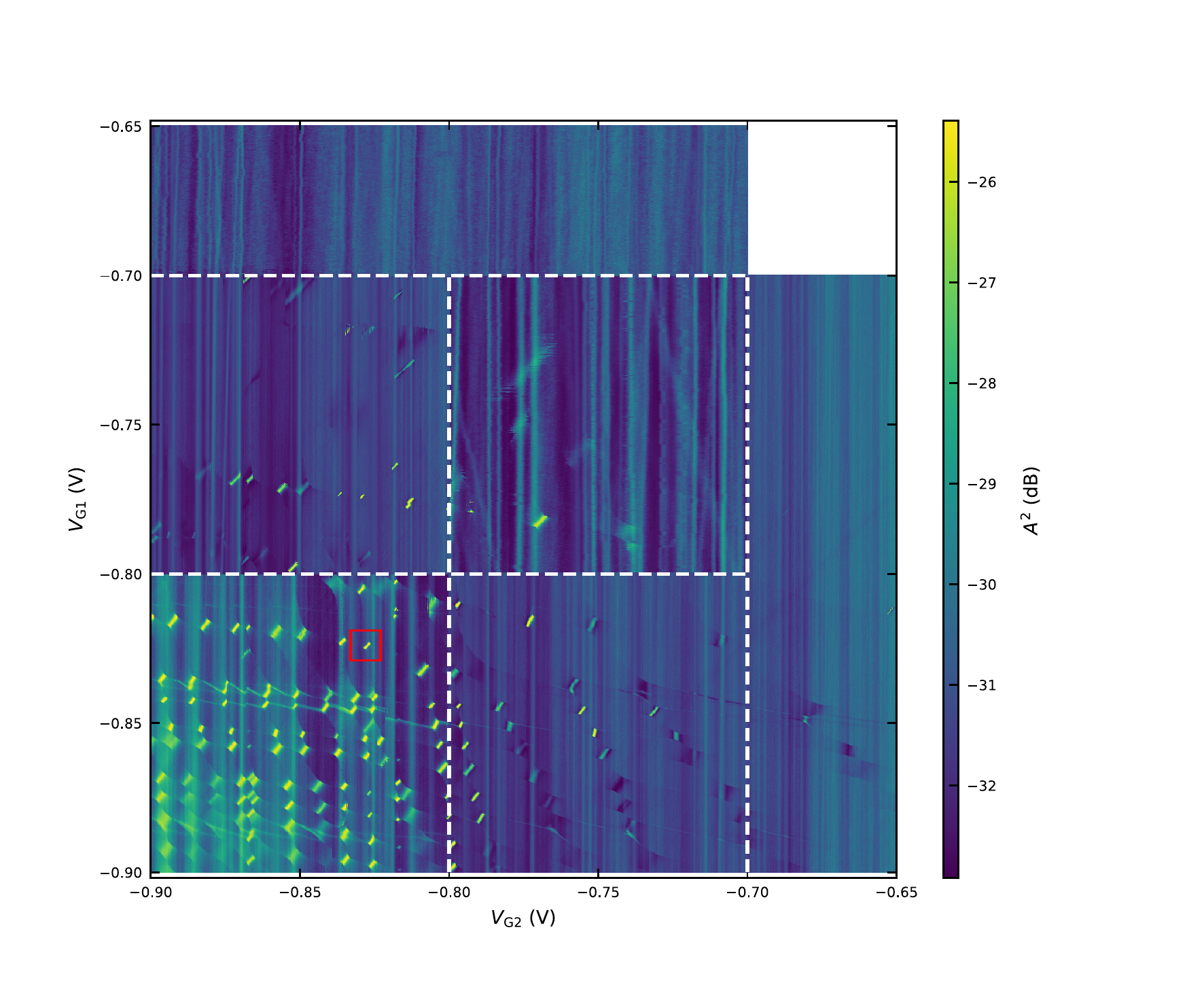}
	\caption{\label{fig:stability_overview}\textbf{Charge stability diagram.} Transmission probed at frequency $\omega_\mathrm{r}/2\pi$ as a function of $\gone$ and $\gtwo$. Several individual measurements (separated by white dashed lines) are stitched together. The interdot charge transition investigated in this work is marked by a red box.}
\end{figure*}

Fig.~\ref{fig:stability_overview} shows a charge stability map with respect to gates G1 and G2 at $\vsd=\SI{0}{V}$ and $\gthree=\SI{0}{V}$. The transmission is probed at the resonance frequency $\omega_\mathrm{r}/2\pi$ of the bare cavity. The interdot charge transitions hence appear as peaks in this map. Indeed, the transmission increases when the electric field of the resonator can move a hole back and forth between the dots, which leads to a dispersive shift of the resonance frequency (see section \ref{sec:intro}). The interdot charge transition investigated in detail in the Main Text is highlighted by a red box. This measurement confirms that the working point is in the few holes regime, with a low number of holes in the QD below G1 ($\sim 4$) and G2 ($\sim 7$).

\section{\label{sec:intro}Charge-photon coupling characterization}
We consider a resonator with angular resonance frequency $\omega_\mathrm{r}$, interacting with a two-level system, such as a charge qubit, with angular frequency $\omega_\mathrm{c}$. This interaction leads to a dispersive shift of the cavity resonance, $\tilde{\omega}_\mathrm{r
} = \omega_\mathrm{r}+\chi_\mathrm{c}$, where $\chi_\mathrm{c}$ is given by \cite{2021_Blais}
\begin{equation}
\label{eq:resonator_qubit}
    \chi_\mathrm{c}=\gc^2d_{01}^2(p_0-p_1)\left(\frac{1}{\Delta}+\frac{1}{\omega_\mathrm{c}+\omega_\mathrm{r}}\right)\,.
\end{equation}
Here $\gc$ is the charge-photon coupling strength, $p_{0,1}$ are the populations of the ground- and excited states of the two-level system, and $d_{01}$ is the dipole moment associated with the transition from the ground to the excited states. The term $g^2_c/\Delta$, with $\Delta=\omega_\mathrm{c}-\omega_\mathrm{r}$, is the Lamb shift, and the counter-rotating term $\gc^2/(\omega_\mathrm{s}+\omega_\mathrm{r})$ is known as the Bloch-Siegert shift. The latter is particularly important in the strong dispersive and ultrastrong coupling regimes~\citep{2021_Blais, 2019_Forn-Diaz, 2019_Kockum}.

The transition frequency and dipole moment associated to a charge qubit are
\begin{subequations}
\label{eq:charge_qubit}
\begin{align}
    \label{eq:charge_qubit_energy}
    \hbar\omega_\mathrm{c}&=\sqrt{\varepsilon^2+4t_\mathrm{c}^2}\,, \\
    \label{eq:charge_qubit_dipole}
    d_\mathrm{01}&=\frac{2t_\mathrm{c}}{\sqrt{\varepsilon^2+4t^2_\mathrm{c}}}\,,
\end{align}
\end{subequations}
with $t_\mathrm{c}$ the tunnel coupling energy, and $\varepsilon$ the energy detuning between the two quantum dots. Hence, at zero detuning ($\varepsilon=0$), the dispersive shift reads
\begin{equation}
	\label{eq:chi}
	\chi_\mathrm{c} = 2\gc^2\frac{\omega_\mathrm{c}}{\omega_\mathrm{c}^2 - \omega_\mathrm{r}^2}(p_0-p_1), 
\end{equation}
with the thermal occupation probabilities
\begin{subequations}
\label{eq:probabilities}
\begin{align}
    \label{eq:probabilities1}
    p_1 &= \frac{1}{1+ e^{\hbar\omega_\mathrm{c}/(k_\mathrm{B}T)}}\,,\\
    \label{eq:probabilities2}
    p_0 &= 1-p_1\,,
\end{align}
\end{subequations}
where $k_\mathrm{B}$ is the Boltzmann constant and $\hbar$ the reduced Planck constant.

\begin{figure*}[htbp]
	\includegraphics[width=\textwidth]{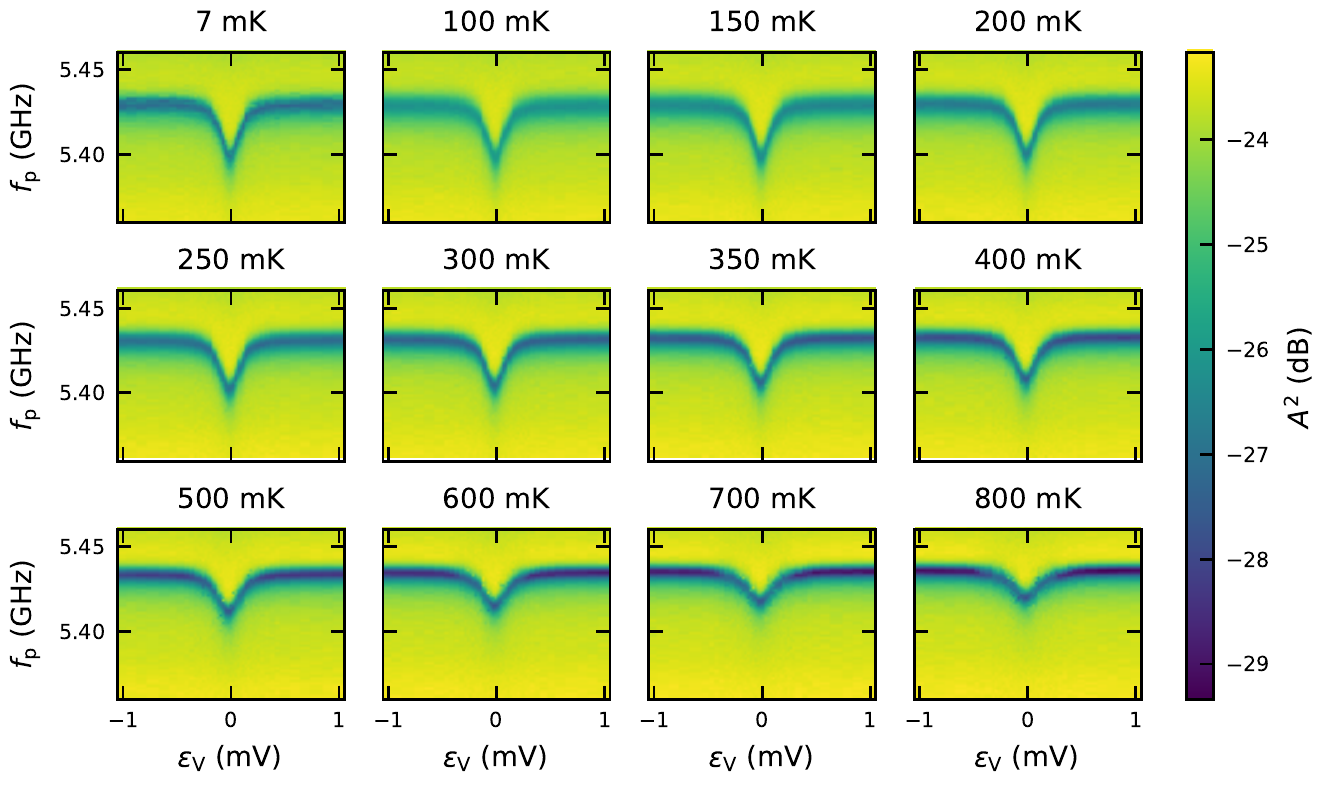}
	\caption{\label{fig:T_dep}\textbf{Temperature dependence of the resonator response.} Transmission as a function of probe frequency $f_\mathrm{p}$ and $\varepsilon$ for various temperatures. At large $|\varepsilon|$ the bare resonator is probed, whereas near $\varepsilon=0$ the charge qubit interacts dispersively with the resonator, which results in a frequency shift. This shift reduces with increasing temperature due to the larger thermal occupation of the excited state of the charge qubit.}
\end{figure*}

\begin{figure*}[htbp]
	\includegraphics[scale=1]{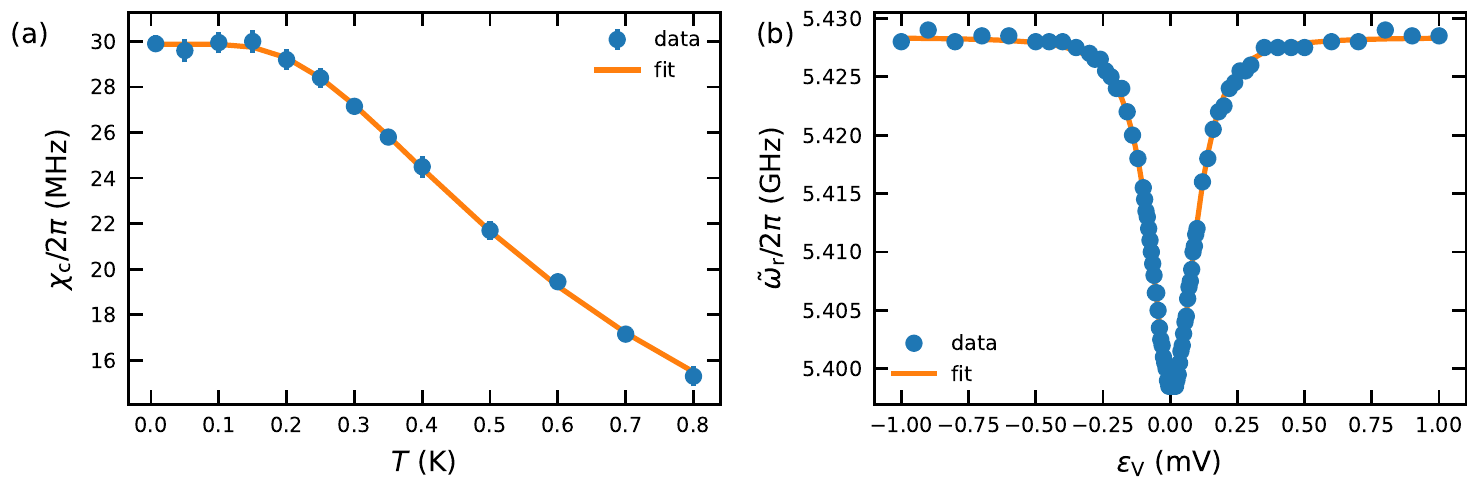}
	\caption{\label{fig:charge_qubit}\textbf{Charge photon coupling and lever arm.} (a) Dispersive shift $\chi_\mathrm{c}$ as a function of temperature extracted from the data shown in Fig.~\ref{fig:T_dep} and fit using Eq.~(\ref{eq:chi}). The error bars represent the standard deviation from the fitting. (b) Resonance frequency at base temperature as a function of $\varepsilon_\mathrm{V}$. Knowing $\gc$ and $t_\mathrm{c}$ from (a), we can extract the lever arm $\alpha$.}
\end{figure*}

Fig.~\ref{fig:T_dep} shows the transmission as a function of voltage detuning $\varepsilon_\mathrm{V}$ and probe frequency $f_\mathrm{p}$ at zero magnetic field and at different temperatures. From these measurements, we extract the dispersive shift $\chi_\mathrm{c}$ at $\varepsilon_\mathrm{V}=0$ and plot it as a function of temperature in Fig.~\ref{fig:charge_qubit}~(a). Fitting the temperature dependence of $\chi_\mathrm{c}$ with Eqs. (\ref{eq:chi}) and (\ref{eq:probabilities}) yields $\gc/2\pi=\SI{513(2)}{MHz}$, $t_\mathrm{c}/h=\SI{9.57(6)}{GHz}$ and $\omega_\mathrm{r}/2\pi=\SI{5.42835(6)}{GHz}$, with $h$ the Planck constant. The fact that the measured $\chi_\mathrm{c}(T)$ is perfectly reproduced by Eqs. (\ref{eq:chi}) and (\ref{eq:probabilities}) confirms the two-level nature of the charge qubit.

We define the lever arm as $\alpha=(1/e)\partial\varepsilon/\partial\varepsilon_\mathrm{V}$, where $\varepsilon_\mathrm{V}=\beta_1\gone-\beta_2\gtwo$ with $\beta_1=0.68$ and $\beta_2=0.73$ is the detuning axis in gate voltage (see black arrow in Fig. 1~(c) of the Main Text). This lever arm can be extracted from the detuning dependence of $\chi_\mathrm{c}$ at base temperature, shown in Fig.~\ref{fig:charge_qubit}~(b). Fitting the data with Eqs.~(\ref{eq:resonator_qubit}), (\ref{eq:charge_qubit_energy}) and (\ref{eq:charge_qubit_dipole}) yields $\alpha=\SI{0.607(3)}{}$. Such large lever arms are commonly measured in these devices and result from the tight electrostatic control by the $\Omega$-shaped gates and the thin gate oxide. 

We can give an alternative estimate of $\gc$ from the lever arm and zero point voltage fluctuation $V_\mathrm{zpf}$ of the resonator. From the design of the bare resonator, we evaluate its impedance $Z_\mathrm{r}\approx\SI{2.5}{\kilo\ohm}$, which results in $V_\mathrm{zpf}=\omega_\mathrm{r}\sqrt{\hbar Z_\mathrm{r}/\pi}\approx\SI{10}{\micro V}$, and in $\gc/2\pi=\alpha\beta_2 eV_\mathrm{zpf}/(2h)\approx\SI{540}{MHz}$. This rough estimate agrees pretty well with the measured value $\gc/2\pi=\SI{513}{MHz}$.

The charge-photon coupling could be further improved by either increasing the impedance of the resonator or by increasing the lever arm. The latter can be enhanced by replacing the SiO$_2$ gate dielectric by a material with a larger dielectric constant or by changing the gate geometry to increase the gate capacitance (e.g. a wrap-around gate).

\section{\label{sec:maps}Detuning-magnetic field maps}

\begin{figure*}[htbp]
	\includegraphics[width=\textwidth]{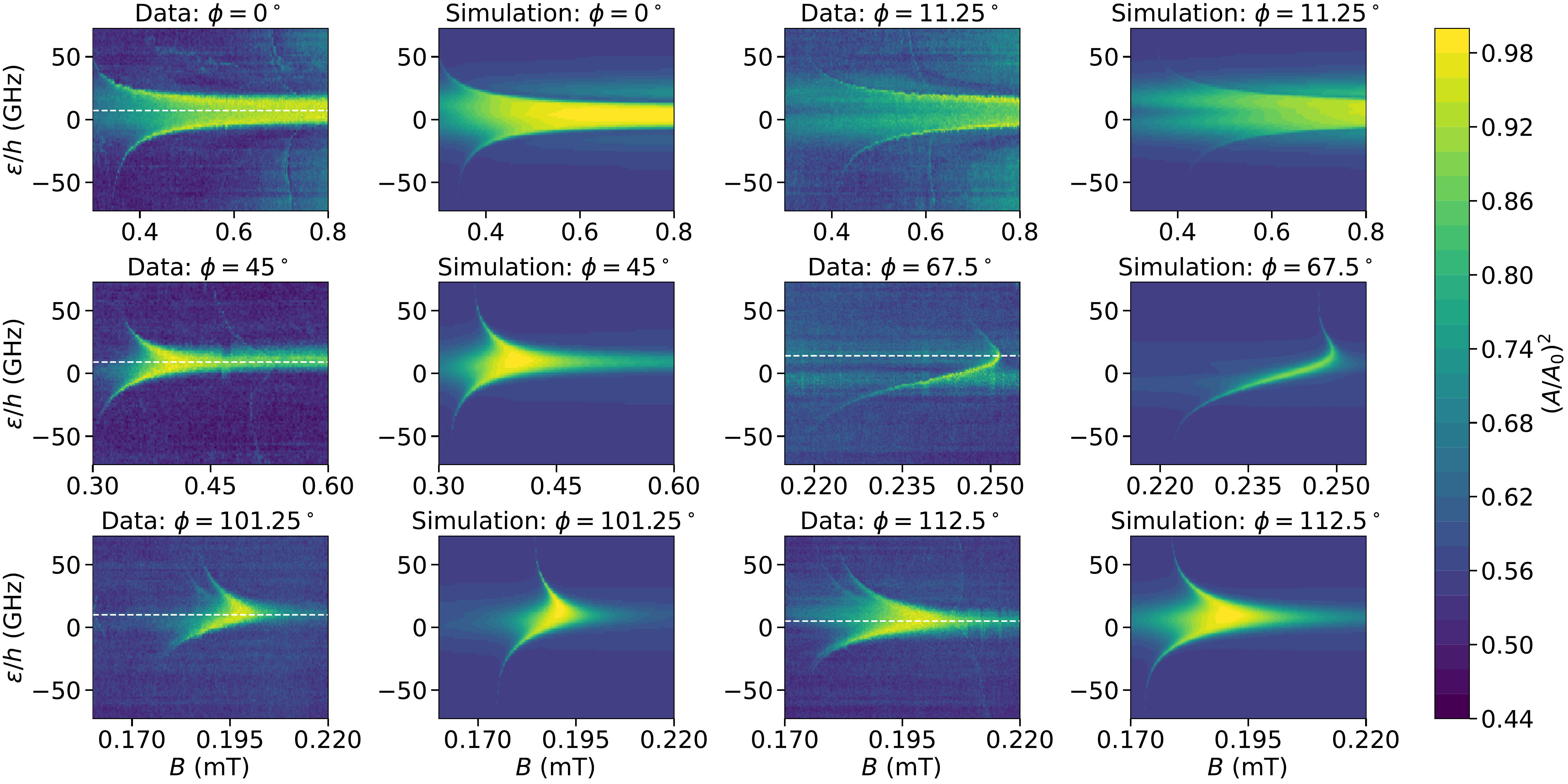}
	\caption{\label{fig:maps}\textbf{Detuning-field maps.} Comparison between the measured and calculated transmission as a function of magnetic field strength $B$ and detuning energy $\varepsilon$, for different magnetic field angles $\phi$. The transmission amplitude $A$ is normalized with respect to the maximum $A_0$ out of the cavity resonance. The simulations are performed with the multi-level input-output theory of Sec. \ref{sec:iotheory}, using the Hamiltonian of section \ref{sec:holemodel} with the parameters of Table \ref{tab:fit}, $\kappa_\mathrm{ext}/2\pi=\SI{3.5}{MHz}$ and $\kappa_\mathrm{int}/2\pi=\SI{10}{MHz}$. The white dashed line is the detuning energy at which the sweet-spot condition $\partial\omega_\mathrm{s}/\partial\varepsilon=0$ is met at resonance $\omega_\mathrm{s}=\omega_\mathrm{r}-\chi_\mathrm{c}$. Namely, the resonance and sweet-spot conditions are simultaneously fulfilled at the maximum of transmission along the white dashed lines. For some angles $\phi$ the sweet spot condition can not be met at resonance, see sections \ref{sec:bres} and \ref{sec:holemodel}.}
\end{figure*}

As a way to characterize the coupled hole-cavity system, it is very useful to map the resonator response as a function of $\varepsilon$ and $B$ for different magnetic field angles $\phi$. For a given set of coordinates $(\varepsilon,B)$, we measure the transmission of a probe tone with angular frequency $\omega_\mathrm{p}(\varepsilon)=\omega_\mathrm{r}-\chi_\mathrm{c}(\varepsilon)$ adjusted to the charge-shifted resonator. In this way, we can track the detuning and field coordinates at which the hole spin is resonant with the cavity, which leads to an increase of the transmission. A few representative detuning-field maps are plotted in Fig.~\ref{fig:maps} for magnetic field orientations that exhibit different degrees of spin-orbit mixing. In particular, we can select in these maps the detuning energy $\varepsilon(\phi)$ at which the spin transition frequency $\omega_\mathrm{s}/2\pi$ satisfies the sweet spot condition $\partial\omega_\mathrm{s}/\partial\varepsilon=0$ at resonance (horizontal white dashed lines in Fig.~\ref{fig:maps}). At such detuning sweet spots, the coupled hole-cavity systems is least sensitive to detuning noise (see section \ref{sec:noise}). All measurements shown in the Main Text have been performed at these sweet spots.

We also compare in Fig.~\ref{fig:maps} the experimental maps with those calculated using the multi-level input-output theory of section~\ref{sec:iotheory} and the Hamiltonian of section~\ref{sec:holemodel}. The simulated maps show very good agreement with the measurements. 

\section{\label{sec:anticrossings}Angular dependence of the strong spin-photon coupling}
\begin{figure*}[htbp]
	\includegraphics[width=\textwidth]{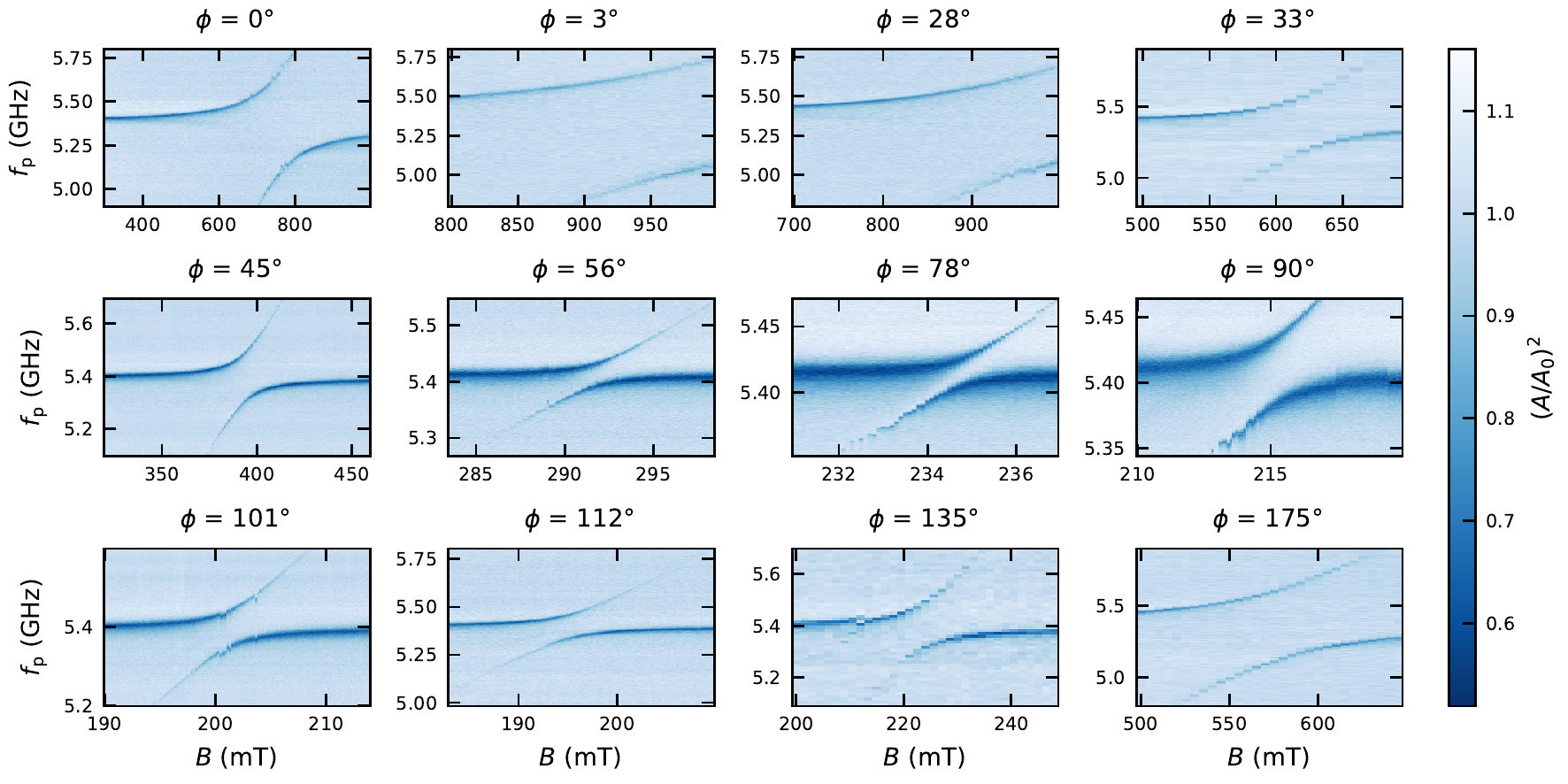}
	\caption{\label{fig:anticrossings}\textbf{Strong spin-photon coupling.} Normalized transmission as a function of probe frequency $f_\mathrm{p}$ and magnetic field strength $B$ for different magnetic field angles $\phi$.}
\end{figure*}
Fig.~\ref{fig:anticrossings} shows the evolution of the avoided crossings as the magnetic field angle $\phi$ is varied. These measurements are acquired at the detuning sweet spot introduced in section~\ref{sec:maps}, which corresponds to the value of $\varepsilon$ indicated by the white dashed line in Fig. \ref{fig:maps}. The vacuum Rabi cuts shown in Fig.~3~(a) of the Main Text are extracted from these measurements.

None of the avoided crossing in Fig.~\ref{fig:anticrossings} does show any sign of other transitions in the Jaynes-Cummings ladder at resonance except the ones involving the resonator ground state. We, therefore, conclude that the cavity is in its ground state and that neither thermal, nor read out driven excitations are significant.

\section{\label{sec:bres}Resonance field}
\begin{figure*}[htbp]
	\includegraphics[scale=1]{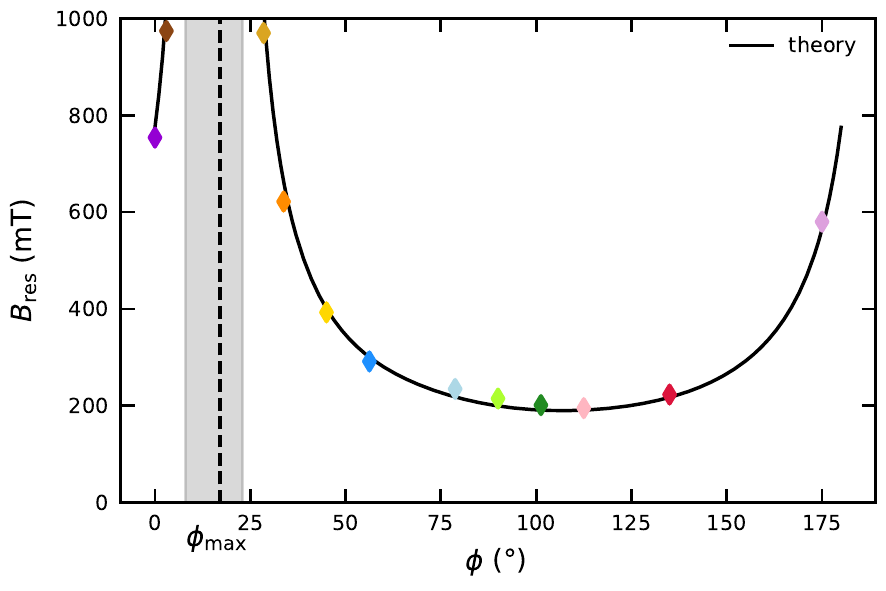}
	\caption{\label{fig:res_field}\textbf{Resonance B field.} Experimental (symbols) and theoretical (line) resonance field as a function of the angle $\phi$ between the $x$ axis and the magnetic field in the $xy$ plane. The shaded gray area outlines the regime where the resonance condition can not be achieved at the detuning sweet spot. The black dashed line indicates the magnetic field angle at which the spin-charge mixing is maximal. The uncertainty of the experimentally measured resonance field is smaller than the marker size.}
\end{figure*}

Fig.~\ref{fig:res_field} shows the magnetic field strength $B_\mathrm{res}$ at which the avoided spin-photon crossing occurs as a function of the magnetic field angle $\phi$. The detuning energy is adjusted to meet the sweet spot condition at resonance (namely, $B_\mathrm{res}$ is measured along the white dashed lines of Fig.~\ref{fig:maps}). The experimental values are compared with the model of section \ref{sec:holemodel}. The data and theory curve are cut above $B_\mathrm{res}=\SI{1}{T}$ for two reasons. First, this range of magnetic field is not accessible experimentally. Second, the resonance and sweet spot conditions can not be met together (see discussion in section \ref{sec:holemodel} for more details).

\section{\label{sec:linewidths}Resonator and spin line widths}
\begin{figure*}[htbp]
	\includegraphics[scale=1]{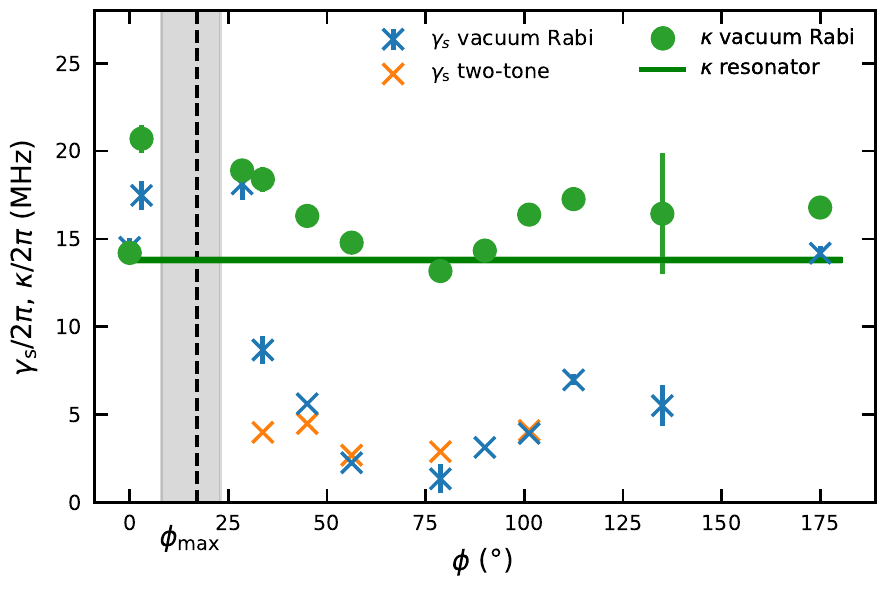}
	\caption{\label{fig:line_widths}\textbf{Angular dependence of $\gamma_\mathrm{s}$ and $\kappa$.} $\kappa$ is extracted from vacuum Rabi mode splittings (green dots) and from the resonance when the number of charges is fixed in the DQD (solid green line, with the light green area indicating the uncertainty, see section \ref{sec:cavity}). $\gamma_\mathrm{s}$ is extracted at the detuning sweet spot from vacuum Rabi mode splittings (blue crosses) and from two-tone spectroscopy far detuned from the spin-photon interaction (orange crosses). The shaded gray area outlines the regime where the resonance condition can not be achieved at the detuning sweet spot. The black dashed line indicates the magnetic field angle at which the spin-charge mixing is maximal. The error bars represent the standard deviation from the fitting.}
\end{figure*}

The spin decoherence rate $\gamma_\mathrm{s}$ and the cavity decay rate $\kappa$ can be drawn from the analysis of the avoided crossings and are plotted as a function of $\phi$ in Fig.~\ref{fig:line_widths}. The bare resonator line width from section \ref{sec:cavity} (with both the spin and charge qubit largely detuned) is shown by the green solid line. The green dots are the $\kappa$'s extracted at finite magnetic field with the spin transition still largely detuned but the charge qubit interacting dispersively with the resonator ($\epsilon$ set to the detuning sweet spot introduced in section~\ref{sec:maps}). Hence, this estimate of $\kappa$ is slightly above the bare cavity loss rate. Knowing $\kappa$, we can infer $\gamma_\mathrm{s}$ from the combined line width $(\gamma_\mathrm{s} + \kappa/2)/2$ obtained by fitting two Lorentzian lines to the vacuum Rabi mode splitting. The uncertainty on the resulting $\gamma_\mathrm{s}$ is however large when the combined line width is fully dominated by $\kappa$. Therefore, we also measure the line width of the spin transition with two-tone spectroscopy for some angles. With the spin largely detuned from the resonator ($|\omega_\mathrm{r}-\omega_\mathrm{s}|\sim 10\gs$), we probe the transmission at the resonator frequency while we sweep a second spectroscopy tone across the spin transition frequency. $\gamma_\mathrm{s}$ is then extracted from the line width of the phase shift of the transmission due to the continuously excited spin transition. The two methods are found to be in good agreement.

\section{\label{sec:holemodel}Flopping mode hole DQD model}

In this section, we discuss the physics of a hole in a double quantum dot.

Holes in the valence band of usual semiconductors such as silicon are subject to a strong spin-orbit (SO) interaction~\cite{1955_Luttinger, 2003_Winkler}. This relativistic effect couples the spin of the hole to its motion in real space. In a semi-classical picture, the ions moving in the rest frame of the hole create a magnetic field that act on the spin attached to this frame.

The valence band wave functions of silicon are essentially linear combinations of the $3p_x$, $3p_y$ and $3p_z$ orbitals of the silicon atoms. The SO interaction thus mixes the angular momentum $l=1$ of these orbitals with the spin $s=1/2$ of the hole. In the present silicon quantum dots, which are most strongly confined along $z=[001]$, the low-energy hole wave functions are a mixture of a majority heavy-hole component (total angular momentum $j=3/2$, $j_z=\pm 3/2$) and a minority light-hole component ($j=3/2$, $j_z=\pm 1/2$) \cite{2018_Kloeffel, 2018_Venitucci, 2021_Michal}. This interplay between different total angular momentum components results in highly anisotropic hole properties. In particular, the gyromagnetic factors of the hole (see below) reflect the heavy-hole/light-hole mixing in the wave function \cite{2016_Voisin, 2021_Liles, 2022_Piot}.

In the following, we consider two ``left'' and ``right'' quantum dots coupled by tunneling across a barrier, in a static, external magnetic field $\mathbf{B}$ (hole double-dot flopping mode qubit \cite{2021_Mutter}). We first discuss the physics and Hamiltonian of this system. We highlight, in particular, the similarities and differences with electron double-dot systems. We next extract the parameters of this Hamiltonian from the experimental data. Finally, we introduce a simple expression for the spin-photon coupling valid in the conditions of the present experiment, and analyze the anisotropy of the spin-photon coupling.

\subsection{Hamiltonian}

We first consider the limit of uncoupled left and right quantum dots. At zero magnetic field, the ground-state levels of each dot are twofold degenerate owing to time-reversal symmetry (Kramers degeneracy).\footnote{This holds for single hole as well as odd fillings of each dot, where the interacting ground-state is generally a Kramers doublet.} They can therefore be mapped onto ``pseudo-spin'' states $\{\ket{L,\Uparrow},\ket{L,\Downarrow}\}$ in the left dot and $\{\ket{R,\Uparrow},\ket{R,\Downarrow}\}$ in the right dot. We emphasize that this mapping is not unique, as any pair of orthogonal eigenstates in the degenerate left dot subspace is a possible choice for $\{\ket{L,\Uparrow},\ket{L,\Downarrow}\}$ (and likewise in the right dot subspace). Owing to the strong mixing between spin and orbital degrees of freedom, the hole wave function cannot, usually, be characterized by its physical spin (as usually done in electron quantum dots). We will come back shortly to the identification of a ``physically meaningful'' pseudo-spin mapping for holes.

We write down the Hamiltonian of the coupled double dot in the $\{\ket{L,\Uparrow},\ket{L,\Downarrow},\ket{R,\Uparrow},\ket{R,\Downarrow}\}$ basis set. For that purpose, we introduce the operators $\tau_L=\ket{L}\bra{L}$ and $\tau_R=\ket{R}\bra{R}$ projecting respectively on the left and right dot subspaces, the Pauli operators $\sigma_x$, $\sigma_y$ and $\sigma_z$ acting in the pseudo-spin subspaces, and the operators $\tau_x=\ket{R}\bra{L}+\ket{L}\bra{R}$, $\tau_y=i(\ket{R}\bra{L}-\ket{L}\bra{R})$ and $\tau_z=\ket{L}\bra{L}-\ket{R}\bra{R}$ \citep{2017_Benito}. The Hamiltonian that describes the detuning energy $\varepsilon$ between the right and left dots then reads for example:
\begin{equation}
  H_\varepsilon=-\frac{\varepsilon}{2}\tau_z\otimes\mathbb{1}_\mathrm{spin}\equiv-\frac{\varepsilon}{2}\tau_z\,,
  \label{eq:Hdetuning}
\end{equation}
where we will generally shorten notations by omitting tensor product symbols and identity operators such as $\mathbb{1}_\mathrm{spin}$ for pseudo-spins. We discuss below the magnetic field and tunneling Hamiltonians.

\subsubsection{Magnetic field and $\gt$-matrices}

To first order,\footnote{The response of the hole becomes non-linear at typical magnetic fields much larger than those reached in the present experiments \cite{2018_Venitucci,2021_Froning}.} the action of the external magnetic field $\mathbf{B}$ on the dots can be generally described by the Zeeman Hamiltonian
\begin{equation}
  H_\mathrm{Zeeman}=\frac{1}{2}\mu_\mathrm{B}\tau_L\left(\boldsymbol{\sigma}\cdot\gt_L\mathbf{B}\right)+\frac{1}{2}\mu_\mathrm{B}\tau_R\left(\boldsymbol{\sigma}\cdot\gt_R\mathbf{B}\right)\,, \label{eq:Hmagnetic}
\end{equation}
where $\mu_\mathrm{B}$ is the Bohr magneton, $\boldsymbol{\sigma}=(\sigma_x,\sigma_y,\sigma_z)$ is the vector collecting the Pauli matrices, and $\gt_i$ is the $3\times 3$ $\gt$-matrix of dot $i=L,R$. Such $\gt$-matrices generalize the isotropic gyromagnetic response of a physical electron spin ($\gt\approx 2\times\mathbb{1}_{3\times3}$) to the possibly anisotropic response of the hole pseudo-spin \cite{1970_Abragam, 2018_Crippa, 2018_Venitucci}. The magnetic field breaks time-reversal symmetry and lifts the Kramers degeneracy in each dot. From Eq.~(\ref{eq:Hmagnetic}), the Zeeman splitting in dot $i$ is $E_\mathrm{Z}=\mu_\mathrm{B}\left|\gt_i\mathbf{B}\right|$ \cite{2018_Crippa}.

As discussed above, the gyromagnetic response of the hole is the result of the interplay between {\it i}) confinement, which shapes an anisotropic hole wave function down to the atomic scale (i.e., a particular mixture of atomic $p_x$, $p_y$ and $p_z$ orbitals), {\it ii}) the SO interaction, which couples the resulting motion of the hole in real space to the physical spin, and {\it iii}) the magnetic field, which acts concurrently on the spin and orbital degrees of freedom \cite{2003_Winkler, 2018_Crippa, 2018_Venitucci, 2022_Piot}. 

It is always possible\footnote{This can be achieved by a singular value decomposition $\gt=U\gp V^\dagger$ \cite{2018_Venitucci}. The diagonal matrix $\gp=\mathrm{diag}(\gt_u,\gt_v,\gt_w)$ defines the principal $\gt$-factors, and the matrix $V$ with columns $\{\boldsymbol{u},\boldsymbol{v},\boldsymbol{w}\}$ defines the principal axes. The real unitary $3\times3$ matrix $U$ can be mapped onto a complex unitary $2\times2$ matrix $U'$ that defines new pseudo-spin states.} to choose the pseudo-spin states $\{\ket{\Uparrow},\ket{\Downarrow}\}$, and a set of orthogonal axes ${\boldsymbol{u},\boldsymbol{v},\boldsymbol{w}}$ for the magnetic field such that the $\gt$-matrix $\gt=\mathrm{diag}(\gt_u,\gt_v,\gt_w)$ is diagonal \cite{2018_Venitucci}. This maps the hole states onto an effective, anisotropic spin with different principal $\gt$-factors $\gt_u$, $\gt_v$, and $\gt_w$ along three principal axes $\boldsymbol{u},\boldsymbol{v},\boldsymbol{w}$. This is actually the pseudo-spin basis where the expression of the Zeeman Hamiltonian is most simple and transparent.

We hence work hereafter in the basis of pseudo-spin states $\{\ket{L,\tilde\Uparrow},\ket{L,\tilde\Downarrow}\}$ and $\{\ket{R,\tilde\Uparrow},\ket{R,\tilde\Downarrow}\}$ that diagonalize the $\gt$-matrix of each dot. We introduce the matrices $\gp_i=\mathrm{ diag}(\gt_u^{(i)},\gt_v^{(i)},\gt_w^{(i)})$ and $V_i=\{\boldsymbol{u}_i,\boldsymbol{v}_i,\boldsymbol{w}_i\}$ collecting the (possibly different) principal $\gt$-factors and axes of dot $i=L,R$. With the magnetic field $\mathbf{B}$ expressed in the original, common device axis set, $\gt_i=\gp_i V_i^\dagger$.

Experimentally, the principal $\gt$-factors and axes can be drawn from the orientational dependence of the Zeeman splitting $E_\mathrm{Z}=\mu_\mathrm{B}\left|\gt\mathbf{B}\right|=\mu_\mathrm{B}[\mathbf{B}\cdot(V\gp^2V^\dagger)\mathbf{B}]^{1/2}$ \cite{2018_Crippa}. They provide valuable information about the heavy-hole/light-hole mixing in the wave function \cite{2022_Piot}.

\subsubsection{Tunneling and Rashba spin-orbit interaction}

The most general expression of the Hamiltonian describing tunneling between the left and right dots, compatible with time-reversal symmetry constraints is:
\begin{equation}
  H_\mathrm{tunnel}=t_0\tau_x-(\mathbf{t}\cdot\boldsymbol{\sigma})\tau_y\,, 
  \label{eq:tunnel}
\end{equation}
where $\mathbf{t}=(t_x, t_y, t_z)$ describes pseudo-spin dependent processes. Note that a change of pseudo-spin states $\{\ket{\Uparrow},\ket{\Downarrow}\}$ in the left or right dot results in a multiplication of $H_\mathrm{tunnel}$ by an unitary matrix, which leaves its determinant invariant. Therefore,
\begin{equation}
  t_\mathrm{c}=\sqrt{t_0^2+|\mathbf{t}|^2}
\end{equation}
is independent on the choice of pseudo-spin states. $t_\mathrm{c}$ is nothing else than the pure ``charge'' tunnel coupling between the dots at zero magnetic field. In the $\{\ket{L,\tilde\Uparrow},\ket{L,\tilde\Downarrow},\ket{R,\tilde\Uparrow},\ket{R,\tilde\Downarrow}\}$ basis set, we introduce a mixing angle $\eta$ and an unit vector $\nso$ such that $t_0=t_\mathrm{c}\cos\eta$ and $\mathbf{t}=t_\mathrm{c}\sin\eta\,\nso$. 

Spin-dependent tunneling may arise because the pseudo-spin states $\ket{\tilde\Uparrow}$ and $\ket{\tilde\Downarrow}$ are different in each dot, and/or because the hole experiences Rashba spin-orbit interactions along the way from one dot to the other. In the former case, a hole with a given spin in the left dot projects onto a different spin in the right dot, which gives rise to a spin-flip process in the tunneling Hamiltonian \cite{2008_Pioro-Ladriere, 2016_Beaudoin}. This mechanism thus becomes ineffective if the two dots are sufficiently similar. On the contrary, Rashba-like spin-orbit interactions act even if the dots are identical. The physics of a long dot or of a DQD tends to be dominated by Rasbha interactions~\cite{2022_Michal}. The Rashba spin-orbit Hamiltonian typically takes the form (for motion along $x$) \citep{2017_Marcellina, 2018_Kloeffel, 2021_Michal, 2021_Froning}:
\begin{equation}
H_\mathrm{so}=\frac{\hbar^2}{m_\parallel\lso}k_x\nso\cdot\,\boldsymbol{\sigma}\,,
\label{eq:SOC}
\end{equation}
where $m_\parallel$ is the effective mass, $\lso$ is the characteristic Rashba spin-orbit length, and $k_x$ is the hole momentum. Equation (\ref{eq:SOC}) can be quite naturally interpreted as a coupling to a spin-dependent gauge field that associates a rotation of the spin around vector $\nso$ to a translation along $x$ (see Refs. \cite{2003_Levitov,2001_Aleiner} for the corresponding unitary transformation of the spin-orbit interaction Hamiltonian for electrons in III-V semiconductors). Then for the double quantum dot the angle of spin rotation is $2\eta=2d/\lso$, where $d$ is the interdot distance. For heavy-holes, $\nso$ is expected to be perpendicular to the direction of motion ($x$) and to the average electric field in the barrier \cite{2021_Michal}, which, for a narrow channel, shall be $z$ -- hence $\nso$ is expected along $y$.

\subsection{Total Hamiltonian and discussion}
\label{sec:Hamanddisc}

In the $\{\ket{L,\tilde\Uparrow},\ket{L,\tilde\Downarrow},\ket{R,\tilde\Uparrow},\ket{R,\tilde\Downarrow}\}$ basis set, the total Hamiltonian of the double dot hence reads:
\begin{align}
  H_\mathrm{DD}&=H_\varepsilon+H_\text{Zeeman}+H_\text{tunnel} \nonumber \\
  &=-\frac{\varepsilon}{2}\tau_z+\frac{1}{2}\mu_\mathrm{B}\tau_L\left(\boldsymbol{\sigma}\cdot\gp_LV_L^\dagger\mathbf{B}\right)+\frac{1}{2}\mu_\mathrm{B}\tau_R\left(\boldsymbol{\sigma}\cdot\gp_RV_R^\dagger\mathbf{B}\right)+t_0\tau_x-\tau_y\left(\mathbf{t}\cdot\boldsymbol{\sigma}\right)\,. 
  \label{eq:toymodel}
\end{align}
This Hamiltonian is fully characterized by the principal $\gt$-factors $\gp_i$ and principal axes $V_i$ of each dot, and by the tunnel couplings $t_0=t_\mathrm{c}\cos\eta$ and $\mathbf{t}=t_\mathrm{c}\sin\eta\,\nso$.

\begin{figure}[tbp]
	\includegraphics[width=14cm]{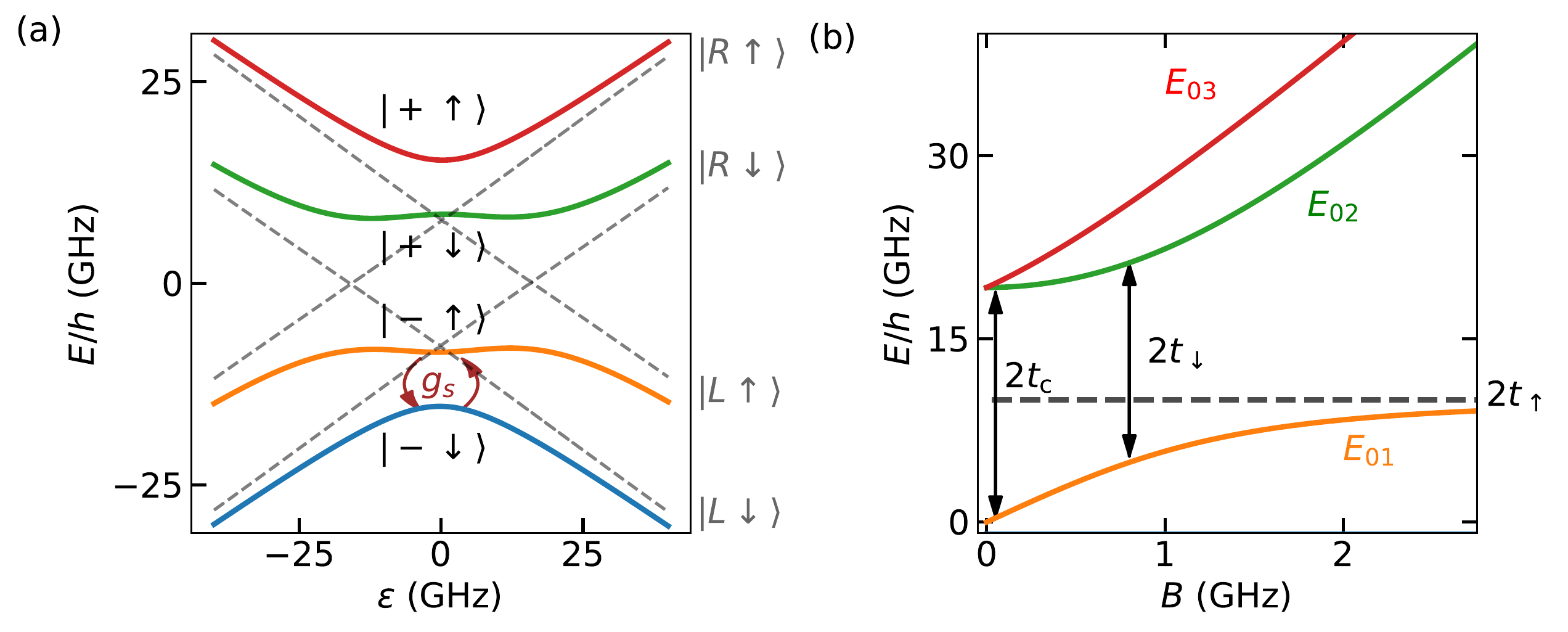}
	\caption{\label{fig:diagram-gs}\textbf{Energy diagram and anticrossings.} (a) Energy diagram as a function of detuning energy $\varepsilon$ for $\phi=\SI{30}{\degree}$, $B=\SI{1}{T}$ and the Hamiltonian parameters of table \ref{tab:fit}. The $\{\ket{L,\uparrow},\ket{L,\downarrow},\ket{R,\uparrow},\ket{R,\downarrow}\}$ states anti-cross due to the tunnel couplings $\tsc$ and $\tsf$. (b) Transition energies from the ground state $\ket{-\downarrow}$ to the excited states as a function of magnetic field strength $B$ for $\varepsilon=0$. The lowest-lying transition from $\ket{-\downarrow}$ to $\ket{-\uparrow}$ is a spin-like excitation whose energy goes to zero when $B\to 0$. This transition acquires an increasing charge-like character when $B$ increases.}
\end{figure}

In the present experiment, the magnetic field $\mathbf{B}=B(\cos\phi,\sin\phi,0)$ lies in the $xy$ plane. For a given magnetic field orientation $\phi$, the Zeeman Hamiltonian $H_\mathrm{Zeeman}$ can be diagonalized. The eigenvectors of $H_\mathrm{Zeeman}$ are the Zeeman-split states of the left dot, labelled $\{\ket{L,\uparrow}(\phi),\ket{L,\downarrow}(\phi)\}$, and the Zeeman-split states of the right dot, labelled $\{\ket{R,\uparrow}(\phi),\ket{R,\downarrow}(\phi)\}$ (with $\downarrow$ the ground-state by convention). We can apply the same unitary transformation $T(\phi)$ that diagonalizes $H_\mathrm{Zeeman}$ to $H_\mathrm{DD}$, and get the total Hamiltonian in the $\phi$-dependent basis set $\{\ket{L,\uparrow},\ket{L,\downarrow},\ket{R,\uparrow},\ket{R,\downarrow}\}$:
\begin{equation}
  H_\mathrm{DD}^\prime(\phi)=T(\phi)^\dagger H_\mathrm{DD}T(\phi)=-\frac{\varepsilon}{2}\tau_z+\tau_L \frac{1}{2}\gt^*_L(\phi)\mu_\mathrm{B}B\sigma_z +\tau_R \frac{1}{2}\gt^*_R(\phi)\mu_\mathrm{B}B\sigma_z + \tsc(\phi)\tau_x - \tsf(\phi)\tau_y\sigma_y, 
  \label{eq:phidep-toymodel}
\end{equation}
where $\gt^*_i(\phi)=|\gp_i V_i^\dagger\mathbf{b}|$ with $\mathbf{b}=(\cos\phi,\sin\phi,0)$ are the effective $\gt$-factors of each dot. This Hamiltonian is the ground for the discussion of the flopping mode hole qubit at a given magnetic field orientation. It gives the following picture of the physics of the device: the Zeeman states in the left and right dot are the eigenstates of the uncoupled dots. They are mixed by inter-dot tunneling, which gives rise to a ``spin conserving'' coupling $\tsc=t_{\downarrow\downarrow}$ and to a ``spin-flip'' coupling $\tsf=-t_{\downarrow\uparrow}$ that respectively preserve and exchange the spin labels of the Zeeman states in the $\phi$-dependent basis set (with $\tsc^2+\tsf^2=t_0^2+\mathbf{t}^2=t_\mathrm{c}^2$). These tunnel couplings result in a family of anti-crossings between $\{\ket{L,\uparrow},\ket{L,\downarrow}\}$ and $\{\ket{R,\uparrow},\ket{R,\downarrow}\}$ as a function of detuning energy $\varepsilon$ for a given value of $\phi$ (see Fig.~\ref{fig:diagram-gs}). In other words, for a given magnetic field orientation, the $\gt$-matrices define a Larmor precession vector $\gt^*_i=\gp_i V_i^\dagger\mathbf{b}$ for the spin in each dot. Due to SO coupling, the tunneling term $H_\mathrm{tunnel}$ gives rise to a competing precession around the axis defined by $\nso$. The interplay between the Larmor and the spin-orbit vectors determines the balance between the effective $\tsc$ and $\tsf$. The expression of $\tsc$ and $\tsf$ as a function of the parameters of Eq.~(\ref{eq:toymodel}) will be given below under particular assumptions relevant for the present experiment.

The Hamiltonian in Eq.~\ref{eq:phidep-toymodel} is formally the same as for the flopping mode electron qubit, but with a set of effective $\gt$-factors and tunnel couplings that highly depend on the magnetic field orientation $\phi$ owing to the strongly anisotropic character of the holes. As a result, the competition between the different terms can be used to engineer the energy levels anti-crossings in Fig.~\ref{fig:diagram-gs}. Note that all parameters of Eq.~(\ref{eq:phidep-toymodel}) are measurable quantities that are independent on the choice of pseudo-spin basis. In particular, $T(\phi)$ can always be chosen such that $\tsc$ and $\tsf$ are real positive. 

We focus on the coupling of photons with the lowest-lying spin transition between the $\ket{-\downarrow}$ and $\ket{-\uparrow}$ branches of Fig.~\ref{fig:diagram-gs}. The coupling strength reads for this transition:
\begin{equation}
\gs=\gc\left|\bra{-\uparrow}\tau_z\ket{-\downarrow}\right|\text{ with }\gc=\frac{1}{2\hbar}\alpha\beta_2 eV_\mathrm{zpf}\,.
\label{eq:gsintro}
\end{equation}
We emphasize that the energy $\hbar\omega_\mathrm{s}$ of this transition, as well as the coupling strength $\gs$ tend to zero when $B\to 0$ owing to time-reversal symmetry, as expected for a ``spin-like'' excitation in a spin-orbit qubit. On the other hand, the $\ket{-\downarrow}\leftrightarrow\ket{-\uparrow}$ transition acquires an increasing ``charge-like'' character with increasing $B$. Indeed, when $B\to\infty$, this transition results from the sole anti-crossing of the $\ket{L,\downarrow}$ and $\ket{R,\downarrow}$ states, so that $\hbar\omega_\mathrm{s}\to 2\tsc$ at zero detuning. 

We next discuss the relation with experimental data and the extraction of the parameters of Eq.~(\ref{eq:toymodel}).

\subsection{Fitting procedure}
\label{sec:Fitting}

Since the magnetic field is only swept in the $xy$ plane, we can not collect enough information to reconstruct all principal $\gt$-factors and axes. In the absence of disorder, strains, and bias asymmetries, the principal axes shall match the device $x$, $y$, $z$ axes for symmetry reasons \cite{2018_Venitucci}. We therefore assume that $z$ remains a (good enough) principal axis of each dot ($\mathbf{w}_L=\mathbf{w}_R=\mathbf{z}$), and look for the two in-plane principal $\gt$-factors $\gp_u^{(i)}$ and $\gp_v^{(i)}$ ($i=L,R$) as well as the two in-plane principal axes $\mathbf{u}_i=(\cos\phi_i,\sin\phi_i,0)$ and $\mathbf{v}_i=(-\sin\phi_i,\cos\phi_i,0)$. Namely, the $\gt$-matrix $\gt_i=\gp_iV_i^\dagger$ of each dot reads:
\begin{equation}
\gt_i=
\begin{bmatrix}
g_u^{(i)}\cos\phi_i & g_u^{(i)}\sin\phi_i & 0 \\
-g_v^{(i)}\sin\phi_i & g_v^{(i)}\cos\phi_i & 0 \\
0 & 0 & g_w^{(i)}
\end{bmatrix}\,,
\end{equation}
where we are only interested in $g_u^{(i)}$, $g_v^{(i)}$, and $\phi_i$. As we achieve an excellent description of the experiment (see below), tilting $\mathbf{w}_L$ and $\mathbf{w}_R$ away from $z$ results in a plain over-parametrization of the model in the absence of out-of-plane data, especially because the in-plane physics is little dependent on a moderate tilt $\lesssim 30^\circ$ \cite{2022_Piot}.

Under this assumption, we can express $\gt^*_i(\phi)$ and $t_{\uparrow\uparrow/\uparrow\downarrow}(\phi)$ as a function of the parameters of Eq.~(\ref{eq:toymodel}). For that purpose, we introduce the following useful quantities:
\begin{subequations}
\label{eq:usefuldefinitions}
\begin{align}
  \gt^*_i&=\sqrt{\left(\gt_u^{(i)}\right)^2+\left(\gt_v^{(i)}\right)^2} \\
  \xi_i&=\arctan\left(\frac{\gt_v^{(i)}}{\gt_u^{(i)}}\right) \\
  \Theta_i&=\arctan\left(\tan(\phi-\phi_i)\tan\xi_i\right) \\
  \Theta_+&=\Theta_L+\Theta_R \\
  \Theta_-&=\Theta_L-\Theta_R\,. 
\end{align}
\end{subequations}
Then,
\begin{subequations}
\label{eq:g0tsctsf}
\begin{align}
  \gt^*_i(\phi)&=\frac{1}{2}\gt^*_i\sqrt{2+2\cos\left(2(\phi-\phi_i)\right)\cos(2\xi_i)} \label{eq:effgfactor} \\
  \tsc^2&=\frac{1}{2}\left(t_\mathrm{c}^2+(t_0^2-t_z^2)\cos\Theta_--2t_0t_z\sin\Theta_-+(t_x^2-t_y^2)\cos\Theta_++2t_xt_y\sin\Theta_+\right) \label{eq:tsc} \\ 
  \tsf^2&=\frac{1}{2}\left(t_\mathrm{c}^2-(t_0^2-t_z^2)\cos\Theta_-+2t_0t_z\sin\Theta_--(t_x^2-t_y^2)\cos\Theta_+-2t_xt_y\sin\Theta_+\right)\,. \label{eq:tsf}
\end{align}
\end{subequations}
Note that $\tsc$ and $\tsf$ may collect SO contributions from $t_0$ and $\mathbf{t}$, from $\gt$-factor anisotropies described by a non-trivial $\Theta_+(\phi)$, and from $\gt$-factor differences through $\Theta_-(\phi)$. The latter term allows for spin-flip tunneling even when $\mathbf{t}=\mathbf{0}$, as the Zeeman-split states in both dots get non-colinear. 

Considering that $t_\mathrm{c}$ and the lever arm $\alpha$ are already known from the charge qubit characterization, we are left with nine parameters in Eqs.~(\ref{eq:g0tsctsf}). These are the six $\gt$-matrix elements $\gt_{u}^{(i)}$, $\gt_{v}^{(i)}$, $\phi_i$, the mixing angle $\eta$, and the angles $\Phi$ and $\Psi$ that define the spin-orbit field $\nso=(\cos\Phi\sin\Psi,\cos\Phi\cos\Psi,\sin\Phi)$.

\begin{table}
\caption{\label{tab:fit} Fitted Hamiltonian parameters (with uncertainties), including both resonator and hole spin terms.}
\begin{centering}
\begin{tabular}{|c|c|c|c|c|c|c|c|c|c|c|c|c|c||}
\hline
   ~$\omega_\mathrm{r}/2\pi$~ & ~$\gc/2\pi$~ & ~$\alpha$~ & ~$t_\mathrm{c}/h$~ & ~$\gt_u^{(L)}$~ & ~$\gt_v^{(L)}$~ & ~$\phi_L$~ & ~$\gt_u^{(R)}$~ & ~$\gt_v^{(R)}$~ & ~$\phi_R$~ & ~$\eta$~ & ~$\Phi$~ & ~$\Psi$~ \tabularnewline
\hline
5.42835 GHz & 513 MHz  & 0.607  & 9.57 GHz & 1.002 & 2.186 & $29.24^\circ$ & 0.922 & 2.248 & $21.03^\circ$ & $83.31^\circ$ & $6.16^\circ$ & $19.75^\circ$ \tabularnewline
\hline
$\pm$0.06 MHz & $\pm$2 MHz  & $\pm$0.004  & $\pm$0.06 GHz & $\pm$0.047 & $\pm$0.078 & $\pm$ $1.18^\circ$ & $\pm$0.037 & $\pm$0.083 & $\pm$ $1.23^\circ$ & $\pm$ $3.06^\circ$ & $\pm$ $2.54^\circ$ & $\pm$ $3.32^\circ$ \tabularnewline
\hline
\end{tabular}
\par\end{centering}
\end{table}

We can relate these parameters to measurable quantities. First, the elements of the $\gt$-matrices of each dot can be obtained from the asymptotic (large $|\varepsilon|$) behavior of the resonance peaks in the detuning-magnetic field maps of the cavity response (see section \ref{sec:maps}). When $|\varepsilon|\gg 2t_\mathrm{c}$, the qubit energy tends to the single-dot limit $\hbar\omega_\mathrm{s}=\mu_\mathrm{B}\gt^*(\phi)B$, which matches the cavity frequency at resonance field $B_\mathrm{res}(\phi)$ such that $\gt^*(\phi)\mu_\mathrm{B}B_\mathrm{res}(\phi)\rightarrow\hbar\omega_\mathrm{r}$. From the so reconstructed $\gt^*_i(\phi)$ and Eq.~(\ref{eq:effgfactor}), we deduce $\gt_u^{(L)}=1.002$, $\gt_v^{(L)}=2.186$, and $\phi_L=29.24^\circ$, as well as $\gt_u^{(R)}=0.922$, $\gt_v^{(R)}=2.248$, and $\phi_R=21.03^\circ$. 

We emphasize though that the $\gt$-factors are expected to depend on bias voltages, and thus on detuning \citep{2013_Ares,2016_Voisin,2019_Studenikin}. The functional form of this dependence is subject to microscopic conditions such as disorder and strain, whose exact modeling goes beyond the scope of this manuscript. However, in the low SO interaction region near $\phi=75^\circ$, the detuning dependence of the $\gt$-factors leads to distinct signatures in the cavity response. We find that the correction $\gt_{u,v}^{(i)}\rightarrow \gt_{u,v}^{(i)}-\delta\gt\exp[-\varepsilon^2/(8t_\mathrm{c}^2)]$, where $\delta\gt=0.06$, yields excellent agreement with the detuning-field maps of section \ref{sec:maps} in this low SO interaction region. This correction is a Gaussian with width $\sigma=2t_\mathrm{c}$, chosen independent on $\phi$ given its negligible effect in the high SO interaction region. We suspect that confinement weakens when the dots slightly delocalize at zero detuning, which leads to a small decrease of the $\gt$-factors. We emphasize that the theory still shows very good agreement with the experiment without this correction. 

To fit the remaining parameters, we make use of several measurements: the detuning-field maps again, the resonance field $B_\mathrm{res}$ at the sweet spot (Fig. \ref{fig:res_field}), the dependence of $\gs$ on $\phi$ at the sweet spot (Fig. 3 of the Main Text), and the dependence of $\gs$ on $\varepsilon$ (Fig. 4 of the Main Text). All these quantities can be calculated numerically from the Hamiltonian, Eq.~(\ref{eq:toymodel}). For those quantities that are measured at the detuning sweet spot, $\varepsilon$ and $B$ are chosen in order to fulfill the resonance ($\omega_\mathrm{s}=\omega_\mathrm{r}-\chi_\mathrm{c}$) and sweet spot conditions ($\partial\omega_\mathrm{s}/\partial\varepsilon=0$)\footnote{The sweet spot is not at $\varepsilon=0$ if the $\gt$-factors of the dots are different.}. The best fit yields $\eta=83.31^\circ$, $\Phi=6.16^\circ$, and $\Psi=19.75^\circ$. 
As discussed in section \ref{sec:Hamanddisc}, the spin transition becomes increasingly charge-like when increasing magnetic field so that its energy $\hbar\omega_\mathrm{s}$ is upper-bounded by $2\tsc$ at zero detuning. Therefore, it is impossible to meet both the resonance and the sweet spot conditions when $2\tsc\lesssim\hbar\omega_\mathrm{r}$. In the present device, this happens in a narrow region of field orientations near $\phi_\mathrm{max}=\SI{17}{\degree}$, outlined in gray on Figs.~\ref{fig:res_field}, \ref{fig:line_widths}, and \ref{fig:rashbahuete} and in Fig.~3 and 4 of the Main Text.

We collect the fitted parameters of both the cavity and hole Hamiltonians in Table \ref{tab:fit}. The orientation of the spin-orbit field and the $\gt$-factors are consistent with the idea that the hole is more strongly confined along the $z$ direction, and, in-plane, more strongly bound in the $y$ direction, as expected in these nanowires \cite{2022_Piot} (see extended discussion in section \ref{sec:DQDanisotropy}). 

\subsection{Analytical calculation of the spin-photon coupling $\gs$}

In this paragraph, we give analytical expressions for the spin-photon coupling strength $\gs$ when the Zeeman splitting is the same in the two dots. This limit is actually relevant for the present device, where the principal $\gt$-factors and axes are close to each other.

For a given magnetic field orientation $\phi$, we introduce the average Zeeman energy of the two dots, $\bar{E}_\mathrm{Z}=\frac{1}{2}(\gt^*_L+\gt^*_R)\mu_\mathrm{B} B$, and the Zeeman energy difference, $\Delta E_\mathrm{Z}=(\gt^*_R-\gt^*_L)\mu_\mathrm{B} B$. We assume $\Delta E_\mathrm{Z}\ll \bar{E}_\mathrm{Z}$, and expand $\gs$ in powers of $\Delta E_\mathrm{Z}/\bar{E}_\mathrm{Z}$. The zero-th order term can thus be obtained as the limit $\Delta E_\mathrm{Z}\to 0$.

Following a similar reasoning as in Ref. \onlinecite{2017_Benito}, we first rewrite the total Hamiltonian, Eq.~(\ref{eq:phidep-toymodel}), in the eigenbasis at $\tsf=0$:
\begin{equation}
    H_\mathrm{DD}^{\prime\prime}=\begin{pmatrix}
    -\frac{1}{2}\left(E_0+\bar{E}_\mathrm{Z}\right) & 0 & 0 & -\tsf \\
    0 & \frac{1}{2}\left(E_0-\bar{E}_\mathrm{Z}\right) & \tsf & 0 \\
    0 & \tsf & -\frac{1}{2}\left(E_0-\bar{E}_\mathrm{Z}\right) & 0 \\
    -\tsf & 0 & 0 & \frac{1}{2}\left(E_0+\bar{E}_\mathrm{Z}\right)
    \end{pmatrix}\,,
    \label{eq:H1}
\end{equation}
where $E_0=\sqrt{\varepsilon^2+4\tsc^2}$ is the spin-conserving band gap. Notice that there are two independent blocks. For compactness, we introduce the left/right mixing angle $\varphi_0$ such that $E_0\cos\varphi_0=2\tsc$ and $E_0\sin\varphi_0=\varepsilon$, as well as the spin-mixing energies and angles $E_+\cos\varphi_+=E_0+\bar{E}_\mathrm{Z}$, $E_+\sin\varphi_+=2\tsf$, $E_-\cos\varphi_-=E_0-\bar{E}_\mathrm{Z}$, and $E_-\sin\varphi_-=2\tsf$. $H_\mathrm{DD}^{\prime\prime}$ is then diagonalized by a rotation of each block by either $\varphi_-$ or $\varphi_+$. The resulting eigenenergies are
\begin{equation}
    \left\{-\frac{E_+}{2}, -\frac{E_-}{2}, \frac{E_-}{2}, \frac{E_+}{2}\right\}, \label{eq:H2}
\end{equation}
with $E_{+}=\sqrt{(E_0+\bar{E}_\mathrm{Z})^2+4\tsf^2}$, and $E_{-}=\text{sign}(E_0-\bar{E}_\mathrm{Z})\sqrt{(E_0-\bar{E}_\mathrm{Z})^2+4\tsf^2}$. We next apply the same rotations to $\tau_z$ in order to get the dipolar matrix elements between the ground-state and the first two excited states:
\begin{subequations}
\begin{align}
    d_{01}&=\cos\varphi_0\sin\left(\frac{\varphi_+-\varphi_-}{2}\right) \nonumber \\
    d_{02}&=\cos\varphi_0\cos\left(\frac{\varphi_+-\varphi_-}{2}\right). \label{eq:dipolar}
\end{align}
\end{subequations}
Depending on the sign of $E_0-\bar{E}_\mathrm{Z}$, the lowest-lying excitation is either the first one with energy $-E_-/2$, or the second one with energy $+E_-/2$. Hence the spin-photon coupling is:
\begin{align}
    \gs&=\gc\left|\theta(E_0-\bar{E}_\mathrm{Z})d_{01}+\theta(\bar{E}_\mathrm{Z}-E_0)d_{02}\right| \nonumber \\
    &=\gc\left|\cos\varphi_0\right|\left|\theta(E_0-\bar{E}_\mathrm{Z})\sin\left(\frac{\varphi_+-\varphi_-}{2}\right)+\theta(\bar{E}_\mathrm{Z}-E_0)\cos\left(\frac{\varphi_+-\varphi_-}{2}\right)\right|,
    \label{eq:gsanalytical}
\end{align}
where $\theta(x)$ is the Heavyside step function. 

This expression reproduces very well the numerical results of Fig.~3~(b) of the Main Text as the $\gt$-factors of the two dots are reasonably close in the present experiment (see Table \ref{tab:fit} and Fig.~\ref{fig:rashbahuete}~(a)). Also, although this approximation neglects Zeeman energy differences in the magnetic Hamiltonian, it still fully accounts for the effect of $\gt$-matrix differences on $\tsf$ (as the latter is computed non-perturbatively with Eq.~(\ref{eq:tsf})).

An insightful limit can be drawn at zero detuning when $\bar{E}_\mathrm{Z}< 2\tsc$ and $\bar{E}_\mathrm{Z}\ll 2t_\mathrm{c}$. Then the spin-charge hybridization is relatively small and Eq.~(\ref{eq:gsanalytical}) reads at leading order in $\bar{E}_\mathrm{Z}/2t_\mathrm{c}$:
\begin{equation}
\gs=\gc\frac{\bar{E}_\mathrm{Z} \tsf}{2t_\mathrm{c}^2}\,.
\label{eq:gsinlowsoc1}
\end{equation}
Moreover, if the $\gt$-matrices $\gt_L=\gt_R=\gt$ are equal ($\gt^*_L(\phi)=\gt^*_R(\phi)$ whatever $\phi$), the Larmor vector $\boldsymbol{\omega}_\mathrm{l}=\mu_\mathrm{B}\gt\mathbf{B}/\hbar$ defines a common quantization axis for the $\ket{\uparrow}$ and $\ket{\downarrow}$ states of both dots. The transformation from Eq.~(\ref{eq:toymodel}) to Eq.~(\ref{eq:phidep-toymodel}) then yields:
\begin{equation}
\tsf=t_\mathrm{c}\sin\eta\left|\mathbf{n}_\mathrm{l}\times\nso\right|\,,
\end{equation}
where $\mathbf{n}_\mathrm{l}=\boldsymbol{\omega}_\mathrm{l}/|\boldsymbol{\omega}_\mathrm{l}|$ is the unit vector along $\boldsymbol{\omega}_\mathrm{l}$. This expression has a simple geometric interpretation: the spin-flip $\tsf$ is proportional to the component of $\mathbf{t}$ in Eq.~(\ref{eq:tunnel}) that is perpendicular to the Larmor vector. We can next introduce an effective spin-orbit field\footnote{Although both $\gt$ and $\nso$ depend on a choice of pseudo-spin basis set, the quantities $g_\mathrm{s}$ and $\mathbf{B}_\mathrm{so}$ do not. In particular, a change of pseudo-spin basis set results in a multiplication of $\gt$ by a real unitary matrix \cite{2018_Venitucci}, which leaves the norm of the cross product invariant.} $\mathbf{B}_\mathrm{so}$ such that $\mu_B\gt\mathbf{B}_\mathrm{so}=t_\mathrm{c}\sin\eta\nso$, and write Eq.~(\ref{eq:gsinlowsoc1}) as:
\begin{equation}
\gs=\gc\frac{\mu_\mathrm{B}^2}{2t_\mathrm{c}^2}\left|(\gt\mathbf{B})\times(\gt\mathbf{B}_\mathrm{so})\right|=\gc\frac{\mu_\mathrm{B}^2}{2t_\mathrm{c}^2}\left|\mathrm{cof}(\gt)(\mathbf{B}\times\mathbf{B}_\mathrm{so})\right|\,,
\label{eq:gsinlowsoc}
\end{equation}
where $\mathrm{cof}(\gt)=\mathrm{det}(\gt){^t}\gt^{-1}$ is the cofactor matrix of $\gt$. The spin-photon coupling is therefore zero when $\mathbf{B}$ is parallel to $\mathbf{B}_\mathrm{so}$. Also note that $\gs\propto B$, as expected for a spin-like transition in a spin-orbit qubit. 

\subsection{Discussion: anisotropy of the DQD properties}
\label{sec:DQDanisotropy}

We conclude with a brief discussion of the angular dependence of the $\gt$-factors and tunnel matrix elements defined by Eqs. (\ref{eq:g0tsctsf}), and of its implications for the physics of the DQD.

\begin{figure*}[htbp]
	\includegraphics[width=\textwidth]{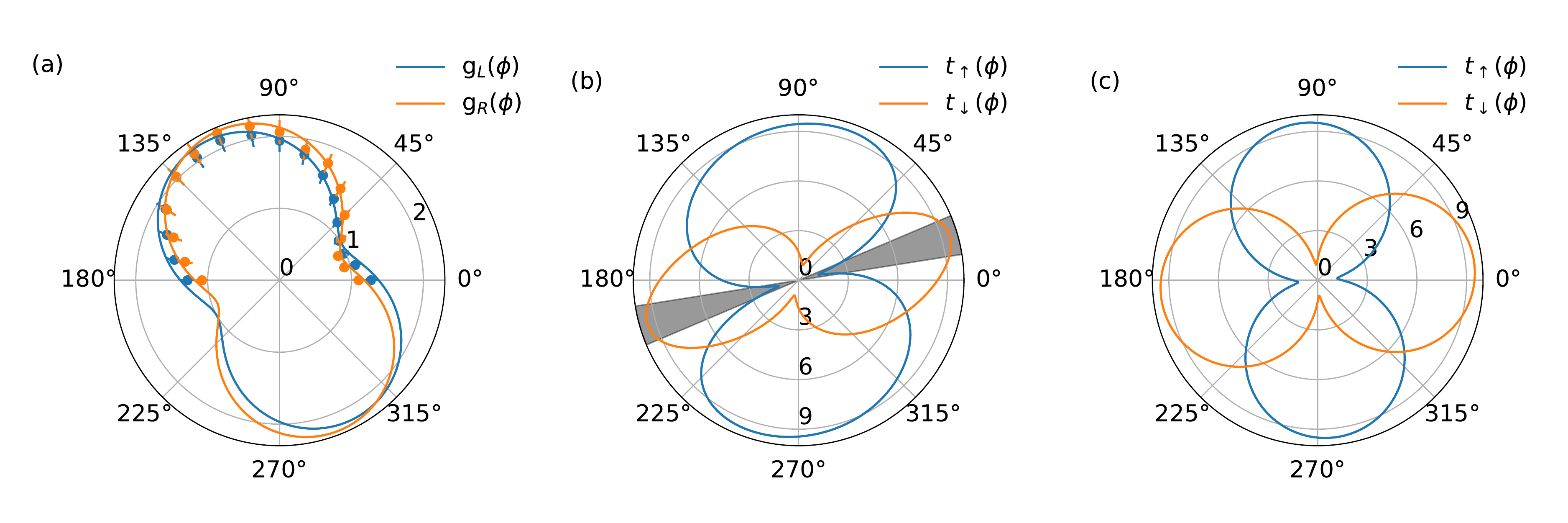}
	\caption{\label{fig:rashbahuete}\textbf{Anisotropy of the $\gt$-factors and tunnel couplings.} (a) In-plane $\gt$-factors of the left and right dots as a function of the angle $\phi$ between the magnetic field and the nanowire axis $x$. The dots are the experimental data and the solid line is the model obtained by fitting the Hamiltonian as described in \ref{sec:Fitting}. The experimental $\gt$-factors are extracted from the resonance condition $\hbar\omega_s=\hbar\omega_r$ at $|\varepsilon|/h=$~\SI{30}{GHz}. Therefore they are a lower bound and come with a systematic uncertainty of $\SI{8}{\percent}$ due to the fact that the wave function still overlaps the two dots at this detuning. (b) Spin-conserving tunnel coupling $\tsc$ and spin-flip tunnel coupling $\tsf$ as a function of $\phi$. The shaded area outlines the angles at which the resonance condition can not be achieved at the sweet spot ($2\tsc\lesssim\hbar\omega_\mathrm{r}$). (c) Expected angular dependence of $\tsc$ and $\tsf$ for an average scalar $\gt$-factor $g_\mathrm{eff}=(g_u^{(L)}+g_u^{(R)}+g_v^{(L)}+g_v^{(R)})/4$.}
\end{figure*}

The in-plane effective $\gt$-factors of the dots are very similar but anisotropic, as shown in Fig.~\ref{fig:rashbahuete}~(a). The $\gt$-factors are indeed much larger along $y$ ($\phi=\SI{90}{\degree}$) than along the channel axis $x$ ($\phi=\SI{0}{\degree}$). This is the signature of a strong heavy-hole/light-hole mixing owing to lateral confinement in the channel \cite{2022_Piot}. Such a strong heavy-hole/light-hole mixing typically enhances spin-orbit interactions \cite{2018_Kloeffel,2018_Venitucci}. Moreover, the principal axes of the $\gt$-matrices are rotated by $\phi_{L,R}\approx\SI{20}{\degree}$ with respect to the device axes. This is most likely the fingerprint of small ($<0.1\%$) process and cool-down shear strains in the device \cite{2022_Piot}. 

The fact that the two dots show similar $\gt$-factors suggests that the dominant spin-flip mechanism is the Rashba SO interaction. In this assumption, the large mixing angle $\eta=\SI{83}{\degree}$ further implies that the spin-orbit length $\lso$ is comparable to the interdot distance $d=\SI{80}{nm}$, which leads to a generally large $\tsf$ as compared to $\tsc$. The angles $\Phi$ and $\Psi$ of Table \ref{tab:fit} define a spin-orbit field $\nso$ lying near the $xy$ plane ($\Phi\approx 0$) and rotated by $\approx\SI{70}{\degree}$ from the nanowire axis $x$ (namely, $\mathbf{B}_\mathrm{so}\approx\nso\approx\mathbf{y}$). 

We plot $\tsc$ and $\tsf$ as a function of $\phi$ in Fig.~\ref{fig:rashbahuete}~(b). The spin-conserving and spin-flip tunnel matrix elements exhibit opposite behaviors, as $\tsc^2+\tsf^2=t_\mathrm{c}^2$. As discussed above, $\tsf\propto|(\gt\mathbf{B})\times\nso|$ to first-order in $B$, so that the spin-charge hybridization is maximal when the Larmor vector and spin-orbit fields are almost perpendicular, namely when $\phi=\phi_\mathrm{max}\simeq\Psi$. Since $\nso$ has a small out-of-plane component ($\Phi\ne 0$), $\tsf$ is nonetheless never zero whatever the finite magnetic field in the $xy$ plane. For comparison, we show in Fig.~\ref{fig:rashbahuete}~(c) the behavior expected for a scalar $\gt$-factor ($\gt\propto\mathbb{1}_{3\times3}$ as for electrons). In that case, the angular dependences are simple trigonometric functions. They are deformed on Fig. \ref{fig:rashbahuete}~(b) by the anisotropic $\gt$-factors of the hole (which define a different scale for each principal axis), and, to a lesser extent, by the small differences between the $\gt$-factors of the dots.

\section{\label{sec:additional_data_single_dot}$\gs$ in the single dot limit}


In the flopping mode model, the value of the spin-photon coupling $\gs$ vanishes rapidly as the DQD is biased away from zero detuning towards the single-dot regime $|\varepsilon|\gg t_\mathrm{c}$ (see Eq. (\ref{eq:gsanalytical})). 

To characterize the value of this single dot coupling, we fit $\gs$ as a function of detuning with:
\begin{equation}
    \gs(\varepsilon)=\gs^{(\mathrm{DD})}(\varepsilon)+p_L(\varepsilon) \gs^{(L)}+p_R(\varepsilon) \gs^{(R)}\,, \label{eq:gssingledot}
\end{equation}
where $p_{L,R}$ are the ground state probabilities of being in left or right dot, $\gs^{(\mathrm{DD})}$ is the spin-photon coupling from the flopping mode DQD model, and $\gs^{(L,R)}$ are the asymptotic spin-photon couplings in the left and right dots. The latter are expected to show some dependence on the orientation of the magnetic field~\citep{2022_Michal}. For $\phi=\SI{11.25}{\degree}$ (Fig. 4 of the Main Text), we extract $\gs^{(R)}/2\pi=\SI{1.16}{MHz}$, and $\gs^{(L)}/2\pi=\SI{0.66}{MHz}$. 


\section{\label{sec:noise}Noise estimation}
Due to the SO interaction, electrical fluctuations perturb the dynamics of the hole spin. Charge noise was shown to be the dominant decoherence mechanism for hole spins even at sweet spots~\citep{2022_Piot}, other decoherence sources such as hyperfine interactions and phonon-induced relaxation being much less effective in comparison. We therefore assume that fluctuations of the detuning energy $\varepsilon$ due to charge noise are the main dephasing mechanism in the DQD. 

We split the spin decoherence rate as $\gamma_\mathrm{s}=\gamma_\varepsilon+\gamma_0$, where $\gamma_\varepsilon$ is due to detuning noise, and $\gamma_0$ accounts for other possible mechanisms, such as hyperfine noise~\citep{fischer2008spin}, electrical $\gt$-factors fluctuations \citep{2022_Piot, 2022_Michal}, or phonon relaxation~\citep{li2020hole}.

We follow Refs.~\citep{2016_Russ, 2020_Koski} to estimate $\gamma_\varepsilon$. For detuning fluctuations $\delta\varepsilon$ with standard deviation $\sigma_\varepsilon$, $\gamma_\varepsilon$ reads to second order:
\begin{equation}
    \frac{\gamma_\varepsilon}{2\pi}=\frac{1}{4\pi^2h}\sqrt{\frac{1}{2}\left(\frac{\partial\omega_\mathrm{s}}{\partial\varepsilon}\right)^2\sigma_\varepsilon^2+\frac{1}{4}\left(\frac{\partial^2\omega_\mathrm{s}}{\partial\varepsilon^2}\right)^2\sigma_\varepsilon^4}\,.
    \label{eq:secondordernoise}
\end{equation}
Considering that the spin decoherence rate $\gamma_\mathrm{s}$ is measured at detuning sweet spots $\partial\omega_\mathrm{s}/\partial\varepsilon=0$, we are left with the second term:
\begin{equation}
    \frac{\gamma_\varepsilon}{2\pi}=\frac{1}{8\pi^2h}\left|\frac{\partial^2\omega_\mathrm{s}}{\partial\varepsilon^2}\right|\sigma_\varepsilon^2\,. 
    \label{eq:sssecondordernoise}
\end{equation}
We next fit $\gamma_0$ and $\sigma_\varepsilon$ on the measured $2\gs/(\gamma_\mathrm{s}+\kappa/2)$ (Fig. 3c of the Main Text), with $\partial^2\omega_\mathrm{s}/\partial\varepsilon^2$ at the sweet spot given by the model of section \ref{sec:holemodel}. We assume the non-dominant term $\gamma_0$ is independent on $\phi$ for simplicity, and use $\kappa/2\approx 2\pi\times 9$ MHz as an average (see Fig.~\ref{fig:line_widths}). We obtain that way $\sigma_\varepsilon=\SI{6.4}{\mu eV}$ and $\gamma_0/2\pi=\SI{3.4}{MHz}$. We note that $\gamma_0$ agrees qualitatively with the data of Fig.~\ref{fig:line_widths}. 

Finally, we can use this fit to estimate the dephasing rate of the charge qubit at the sweet spot $\varepsilon=0$. Assuming $\gamma_0$ only applies to the spin degree of freedom and that detuning noise is again the main decoherence mechanism, we find $\gamma_c/2\pi=\sigma_\varepsilon^2(\partial^2\omega_\mathrm{c}/\partial\varepsilon^2)/(8\pi^2h)=9.9$ MHz for $2t_\mathrm{c}/h=\SI{19.2}{GHz}$. Note that $\partial^2\omega_\mathrm{c}/\partial\varepsilon^2\propto 1/t_\mathrm{c}$, so that a large tunneling gap reduces dephasing. This leads to low dephasing with respect to previous charge-photon coupling experiments \citep{2017_Stockklauser, 2020_Koski, 2021_Kratochwil},  comparable to noise-mitigated charge qubits \cite{2021_Scarlino}. 

\section{\label{sec:iotheory}Multi-level input-output theory beyond the rotating-wave approximation}
Part of the characterization of the hole qubit is based on the measurement of the response of the cavity to a probe field as a function of detuning and magnetic field. Here, we use the input-output (IO) theory to relate the measured quantities with the flopping mode hole DQD model (section \ref{sec:holemodel}). The IO theory is widely used to compute the response of a cavity-spin qubit system to a probe field with amplitude $a_\mathrm{in}$ and angular frequency $\omega_\mathrm{p}$ \citep{2016_Burkard, 2017_Benito}. Considering that our system borders the ultrastrong coupling regime, we need to account for counter-rotating terms that are often neglected in the literature. To go beyond the rotating wave approximation (RWA), we follow Ref.~\onlinecite{2018_Kohler} for cavity readout.

In the DQD eigenbasis at a given working point, the total Hamiltonian of the system is
\begin{equation}
    H/\hbar=\omega_\mathrm{r}a^\dagger a + \omega_n\ket{n}\bra{n} + \gc(a+a^\dagger)d_{nm}\ket{n}\bra{m} \label{eq:totalH}\,,
\end{equation}
where $\omega_n=E_n/\hbar$ with $E_n$ the eigenenergies of the Hamiltonian, Eq.~(\ref{eq:toymodel}), and $d_{nm}$ are the dipolar matrix elements in the DQD eigenbasis.

The Langevin equation for the cavity field $a$ is
\begin{equation}
\label{eq:langevin}
\dot a = -i\omega_\mathrm{r} a -i\gc d_{nm}\ket{n}\bra{m} -\frac{\kappa_\mathrm{ext}+\kappa_\mathrm{int}}{2} a
-\sqrt{\frac{\kappa_\mathrm{ext}}{2}} a_\mathrm{in} \,,
\end{equation}
which is solved to obtain the stationary state $\dot a=0$ using the input-output relation for the hanger geometry $a_\mathrm{in}-a_\mathrm{out}=\sqrt{\kappa_\mathrm{ext}/2}a$~\citep{2015_dumur, 2021_wang}. The transmission coefficient can be estimated assuming that the probe frequency is comparable to the resonator frequency $|\omega_\mathrm{p}-\omega_\mathrm{r}|\ll \omega_\mathrm{r}$, as is the case in the current experiment \citep{2018_Kohler}:
\begin{equation}
    \frac{A}{A_0}\equiv\left|\frac{a_\mathrm{out}}{a_\mathrm{in}}\right|
=\left| 1 + \frac{i\kappa_\mathrm{ext}/2}{\omega_\mathrm{r}-\omega_\mathrm{p}+\chi(\omega_\mathrm{p})-i(\kappa_\mathrm{ext}+\kappa_\mathrm{int})/2}\right|\,,
\label{eq:rc}
\end{equation}
where $\chi(\omega_\mathrm{p})$ is the susceptibility, or angular frequency shift, in the linear response approximation:
\begin{equation}
\chi(\omega_\mathrm{p}) = \gc^2\sum_{m,n}
\frac{(p_n-p_m)|d_{nm}|^2}{\omega_\mathrm{p} -(\omega_m-\omega_n) +i\gamma_{mn}}\,.
\label{eq:susceptibility}
\end{equation}
Here $p_n$ is the probability of eigenstate $n$, and we introduce $\gamma_{mn}$ phenomenologically as the decoherence rate associated to the transition $m\leftrightarrow n$. Note that, in the charge qubit case, we recover the angular frequency shift $\chi=\chi_\mathrm{c}$ of Eqs. (\ref{eq:resonator_qubit}) to (\ref{eq:chi}), including the counter-rotating Bloch-Siegert shift.

Using this theory, we obtain the position of the two vacuum Rabi modes induced by the spin-photon interaction as the zeros of the real part of the denominator of Eq.~(\ref{eq:rc}). The equation to solve is thus
\begin{equation}
    \text{Re}\left[\omega_\mathrm{r}-\omega_\mathrm{p}+\chi(\omega_\mathrm{p})\right]=0  \,.
    \label{eq:rabimodecondition}
\end{equation}
Considering that the measurements are made in the low temperature limit, we can take $p_0=1$; neglecting dephasing, the previous equation then leads to
\begin{equation}
    \omega_\mathrm{r}-\omega_\mathrm{p}-\gc^2\sum_n |d_{n0}|^2\left(\frac{1}{\omega_n-\omega_0-\omega_\mathrm{p}}+\frac{1}{\omega_n-\omega_0+\omega_\mathrm{p}}\right)=0\,,
    \label{eq:rabimodecondition+}
\end{equation}
which is solved for near resonant interaction $\omega_\mathrm{r}\approx \omega_1-\omega_0$. Under the RWA with two levels only, we readily get the Rabi peak positions
\begin{equation}
    \omega_\mathrm{p}=\omega_\mathrm{r}\pm\gc d_{01} \,,
    \label{eq:rabipeakRWA}
\end{equation}
and the Rabi splitting $2\gs=2\gc d_{01}$, as assumed in the Main Text. This simple result does not, however, necessarily hold in the multi-level case beyond the RWA, or when $\gs$ is comparable to $\kappa$ and $\gamma$.

To find out whether the above approximation applies in the presence of multiple levels and counter-rotating terms, we solve Eq.~(\ref{eq:rabimodecondition+}) numerically using the Hamiltonian parameters of Table \ref{tab:fit}. We find that the distance between Rabi peaks can be approximated by $2\gs$ with a relative error below $0.5\%$ in all measured cases. The reason why the corrections are minor is the resonance condition $\omega_\mathrm{r}\approx \omega_1-\omega_0$, which makes the resonant spin transition much stronger than the others (higher-lying transitions are far away due to the large tunnel coupling and therefore show large denominators), and much stronger than the counter-rotating terms in Eq.~(\ref{eq:rabimodecondition+}). 

Finally, when $\gs$ is comparable to $\gamma$ and/or $\kappa$, as is the case in the measurements for the single-dot limit $|\varepsilon|\gg t_\mathrm{c}$, the position of the peaks is no longer given by Eq.~(\ref{eq:rabipeakRWA}). In this regime we use Eq.~(\ref{eq:rc}), including the dephasing and cavity decay rates, to relate the distance between peaks and their amplitudes to $\gs$.

\bibliography{literature}

\end{document}